# Quantum Fourier Transform for Image Processing

Ze Yu Zhang, Weibo Gao

Nanyang Technological University

***Abstract*** **-** Quantum information processing and its sub-field, quantum image processing, are fast growing fields as results of advancement in the practicality of quantum mechanics. In this paper, a quantum algorithm which can be used to process information such as one-dimensional time series and two-dimensional images in frequency domain is proposed. The information of interest is encoded into the magnitude of probability amplitude or the coefficient of each basis state. The oracle for filtering works based on post-selection results and its explicit circuit design is presented. This oracle is versatile enough to perform all basic filtering, namely high pass, low pass, band pass, band stop and many other processing techniques. Finally, two novel schemes for transposing matrices are presented in this paper. They use the similar encoding rules but with deliberate choices in terms of selecting basis states. These schemes could be potentially useful for other quantum information processing tasks such as edge detection. The proposed techniques are conducted on quantum simulator. Some of the results are compared with traditional information processing results to verify its correctness and presented in this paper.

## 1 INTRODUCTION

As quantum technology gets maturing, more and more schemes are proposed to exploit how to use this novel way to accomplish the tasks which are previously achieved by traditional approaches, since the quantum approaches allow certain advantages that traditional means can never accomplish. Notably, Shor's factoring algorithm proposed by Peter Shor in 1994 gives exponential speedup for integer number factorization [1]. Another example would be Grover's algorithm for searching a labeled entry in an unstructured database with $O(\sqrt{N})$ speedup compared to its classical counterpart [2]. Inspired by quantum algorithms alike, people have been trying to implement them in real world and have achieved promising results [3][4][5]. In the field of image processing, various quantum-based approaches have also been investigated [6], hence the birth of an exciting field of research Quantum information processing (QIP). However, among the proposed schemes some has rather complex information encoding method [7] while some is only applicable for grey scale images and does not provide explicit circuits for oracles [8], which is arguably the most essential part for the task. The scheme proposed in this paper is comparatively more intuitive to understand and has explicit circuit design for the oracle. Furthermore, as it will be shown the proposed oracle is essentially free to filter any frequency components and hence its versatility.

## 2 RELATED WORKS

### 2.1.1 Quantum Fourier Transform (QFT)

The quantum Fourier transform on an orthonormal basis$|0\rangle,\cdots,|N-1\rangle$is defined to be a linear operator with the following action on the basis states

$$QFT(|x\rangle) = \frac{1}{\sqrt{N}}\sum_{0}^{N-1}\omega_x^k|k\rangle$$

Where $|x\rangle$ is a basis state and $\omega_x = e^{\frac{2\pi xi}{N}}$.

### 2.1.2 Inverse Quantum Fourier Transform (IQFT)

The IQFT on an orthonormal basis$|0\rangle,\cdots,|N-1\rangle$is defined to be a linear operator with the following action on the basis states

$$IQFT(|k\rangle) = \frac{1}{\sqrt{N}}\sum_{0}^{N-1}\omega_k^x|x\rangle$$

Where $|x\rangle$ is a basis state and $\omega_x = e^{-\frac{2\pi ki}{N}}$.

Or equivalently, IQFT in the inverse operation of QFT.

### 2.1.3 Matrix representation

$$QFT = \frac{1}{\sqrt{N}}\begin{bmatrix} 1 & 1 & 1 & 1 & \cdots & 1 \\ 1 & \omega_n & \omega_n^2 & \omega_n^3 & \cdots & \omega_n^{N-1} \\ 1 & \omega_n^2 & \omega_n^4 & \omega_n^6 & \cdots & \omega_n^{2(N-1)} \\ 1 & \omega_n^3 & \omega_n^6 & \omega_n^9 & \cdots & \omega_n^{3(N-1)} \\ \vdots & \vdots & \vdots & \vdots & \ddots & \vdots \\ 1 & \omega_n^{N-1} & \omega_n^{2(N-1)} & \omega_n^{3(N-1)} & \cdots & a_{NN} \end{bmatrix}$$

and

$$IQFT = (\frac{1}{\sqrt{N}}\begin{bmatrix} 1 & 1 & 1 & 1 & \cdots & 1 \\ 1 & \omega_n & \omega_n^2 & \omega_n^3 & \cdots & \omega_n^{N-1} \\ 1 & \omega_n^2 & \omega_n^4 & \omega_n^6 & \cdots & \omega_n^{2(N-1)} \\ 1 & \omega_n^3 & \omega_n^6 & \omega_n^9 & \cdots & \omega_n^{3(N-1)} \\ \vdots & \vdots & \vdots & \vdots & \ddots & \vdots \\ 1 & \omega_n^{N-1} & \omega_n^{2(N-1)} & \omega_n^{3(N-1)} & \cdots & a_{NN} \end{bmatrix})^{-1}$$

Where $n$ is the number of two-level qubits, $N = 2^n$ and $\omega_n = e^{\frac{2\pi i}{N}}$.

It can be observed that QFT has the same matrix form as inverse discrete Fourier Transform (IDFT) and IQFT has the same matrix form as discrete Fourier Transform (DFT) up to a normalization factor. Therefore, one would expect to use QFT/IQFT for similar use as DFT/IDFT.

So far the most optimal QFT algorithms require $O(nlog(n))$ gates to achieve an efficient approximation [9]. In contrast the most commonly practised classical DFT technique fast Fourier transform (FFT) has a time complexity of $O\big(Nlog(N)\big) = O(2^n log(2^n)$. Clearly QFT has an exponential speedup compared to FFT, let alone DFT which has $O(N^2) = O(2^{2n})$ time complexity.

### 2.2 Quantum Logic Gate

Quantum logic gates are a set of gates that can be applied to qubits and perform certain operations, quantum logic gates perform unitary and reversible operations.

### 2.3 Frequency Domain Filtering

In classical digital signal processing (DSP), a filtering mask is applied to mask out the unwanted frequency components in frequency domain.

### 2.4 Signal Processing Through Fourier Transform

Fourier transform has been widely used in classical signal processing field. In particular, for one-dimensional (1-D) discrete digital signal with n components, first apply DFT to the vectorized signal $[x'_0, x'_1, \dots, x'_{n-1}]^T$ ,and apply a filter mask (can be high-pass, low-pass, band-pass, band-stop etc.) according to the requirement. The resultant vector $[X'_0, X'_1, \dots, X'_{n-1}]^T$ contains the frequency components which is symmetrical about $x'_{\frac{n}{2}+1}$ if n is even and symmetrical about $X'_{\frac{n+1}{2}}$ and $X'_{\frac{n+1}{2}+1}$ if n is odd with $X'_0$ being the DC component. The higher frequency components are nearer to the symmetrical axis and lower frequency components are nearer to the top and bottom rows. Normally a so-called FFT-shift technique will be done to move low frequency components to the center of the array for easier observation but it is trivial for our purpose.

## 3 INFORMATION REPRESENTATION

### 3.1 One-Dimensional Time Series

In this representation of choice, information of interest is simply the discrete data points listed in temporal order. The data points considered are real values. Hence, they can be completely described by the probability amplitude of the basis states.

Assume n qubits $q_0$, $q_1$, $q_2$, …, $q_{n-1}$ have been prepared and there is a time series X = [$x_0$, $x_1$, $x_2$, …, $x_{N-1}$] (where N=$2^n$). To begin with, X is to be normalized such that: $X' = aX = a[x'_1, x'_2, \dots, x'_N]$ and ${x'_1}^2 + {x'_2}^2 + \cdots + {x'_N}^2 = 1$

Now $x'_1, x'_2, \dots, x'_N$ can be encoded onto the probability amplitudes of $|0000 \dots 000\rangle$ , $|0000 \dots 001\rangle$ , …, $|1111 \dots 111\rangle$ . A normalization constant is to be taken note to recover X from the measurement results. The relative phase differences of different states are trivial in this case, since the measured statistical results for probability amplitude are not affected by them. It needs to be noticed that when N is not a power of 2, there will be redundant (unused) states whose coefficients are set to 0 and these values are discarded after extracting the meaningful information.

### 3.2 Two-Dimensional Images

Two-dimensional (2-D) images, regardless of RGB images or grey scale images, are represented in computer as matrices with each entry's value as the intensity of some corresponding color channel. First reshape the matrix into a 1-D array and then the same technique to encode 1-D images can be applied to 2-D images as well.

As an example, let the matrix representation of a grey scale image

$$A = \begin{bmatrix} a_{11} & \cdots & a_{1N} \\ \vdots & \ddots & \vdots \\ a_{N1} & \cdots & a_{NN} \end{bmatrix}.$$

After reshaping it into A',

$A' = [a_{11}, a_{12}, \dots, a_{1N}, a_{21}, a_{22}, \dots, a_{2N}, \dots, a_{N1}, a_{N2}, \dots, a_{NN}]$. It is easy to check now 2n qubits are needed to encode normalized array of A'.

As it will be shown later compared to the commonly used 2-D Fast Fourier Transform, this encoding approach is less accurate but still useful to certain extent. It will be further addressed as to how to further improve the situation at the end of the discussion. As for RGB image, red, green and blue color intensity values can be arranged into a 1-D array as well. However, it will be three times larger of the size of the same image in grey scale representation and the number of qubits should increase correspondingly. Consequently, in some cases (for example A) there will be redundant (unused) states (whose coefficients are again set to zero and later discarded) since the number of elements are not a power of 2n, as for more general 2-D images' matrix.

## 3.3 Viability

To prove that it is actually possible to encode arrays with arbitrary normalized values in such a way, it is sufficient to prove the matrix that describes such an operation is unitary since it has been shown that any unitary operation on quantum circuit can be experimentally realized [10].

Proof:
We want to map from a specific state, say, $|000\ldots000\rangle$ to $x'_1\left|000\ldots001\rangle, x'_2\right|000..001\rangle, \ldots, x'_N|111\ldots111\rangle$ , i.e., $\left[x'_1, x'_2, \ldots, x'_N\right]^T = S[1,0,\ldots,0]^T$ where S is supposedly a transformation that satisfies such a relation. Furthermore, S also maps the rest of standard bases to some other bases. To preserve the dimension of the space, after the transformation all the basis vector should be still orthogonal to each other.

$$
\begin{aligned}
S &= S \times I \\
&= S[1,0,\ldots,0]^T + S[0,1,\ldots,0]^T + \cdots \\
&\qquad + S[0,0,\ldots,1]^T \\
&= \left[x'_1, x'_2, \ldots, x'_N\right]^T + v_2 + \cdots + v_N \\
&= \begin{bmatrix} x'_1 & & & \\ \vdots & v_2 & \cdots & v_N \\ x'_N & & & \end{bmatrix}
\end{aligned}
$$

where $v_i$ are the unknown basis vectors after transformation.

It is easy to show that $S^\dagger \times S = I$ since each column vector of S is normalized (i.e. $\langle v_i|v_j\rangle = 1$ when $i = j$ ) and all the column vectors all orthogonal (i.e. $\langle v_i|v_j\rangle = 0$ when $i \neq$ j ). Following that $S^\dagger = S^{-1}$ and $S \times S^\dagger = I$. Therefore, S is unitary and consequently can be implemented as quantum gate.

# 4 THE QUANTUM FILTER ORACLE

## 4.1 General Circuit Design

The circuit is of the following form:

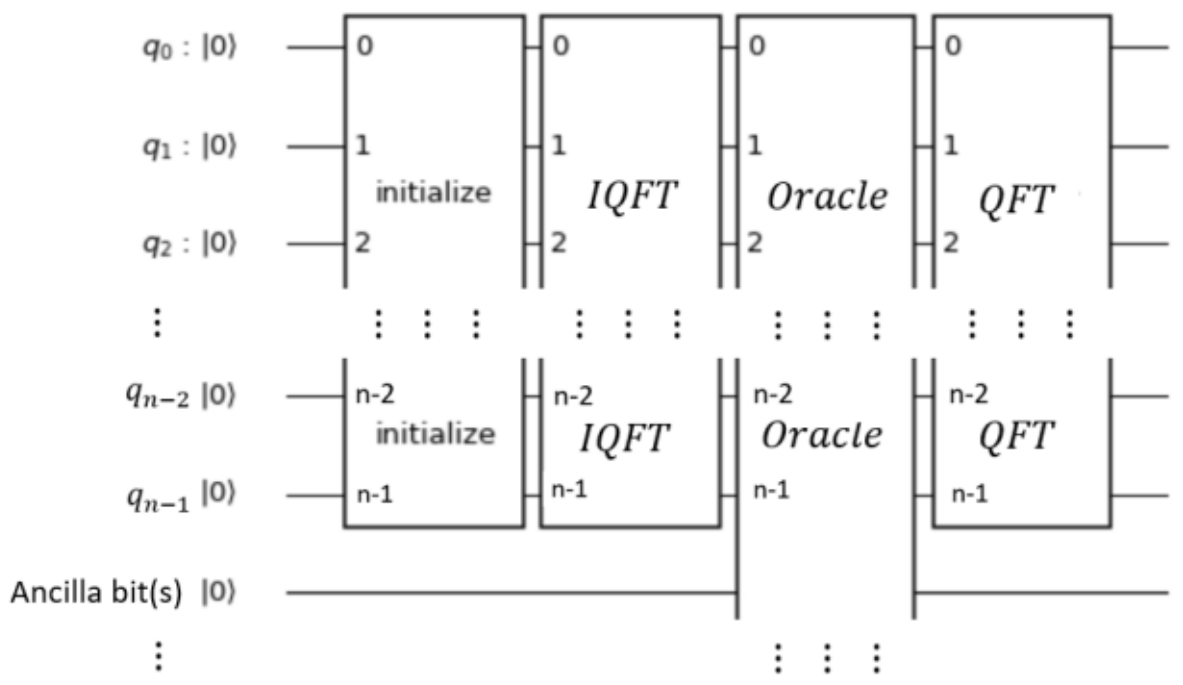

Fig. 1

The first unitary operator encodes normalized values of interest onto the probability amplitudes of the basis states. The second operator is the inverse quantum Fourier transform (IQFT) operator which transforms the time domain to frequency domain. The third operator is an oracle which performs specific filtering in frequency domain. The last operator is the quantum Fourier transform operator (QFT) which transform the frequency domain back to time domain.

The oracle is a postselection subroutine. It marks the unwanted basis states with an ancilla bit, after which the measurements are performed on the ancilla bit. If ancilla bit has a measured result agrees with the unmarked states, it implies the unwanted states have been filtered out and all other states are re-normalized with relative phases unchanged. The number of measurements needs to be performed is related to the total probability amplitudes of the unwanted states. If it is small, then fewer measurements are needed and vice versa.

## 4.2 The Four-Qubit Case (1-D Time Series Processing Simulation on Qiskit)

To illustrate, consider a four-qubit system. First the state of the system is initialized in to $\left[0,0,0,0,0,0,\frac{1}{2},\frac{1}{2},\frac{1}{2},\frac{1}{2},0,0,0,0,0,0,\right]^T$ (the superscript 'T' stands for transpose) as shown in Figure 2.

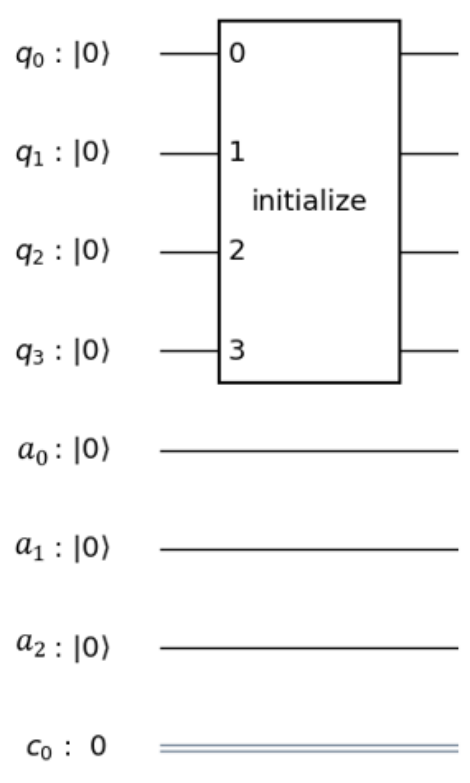

Fig. 2

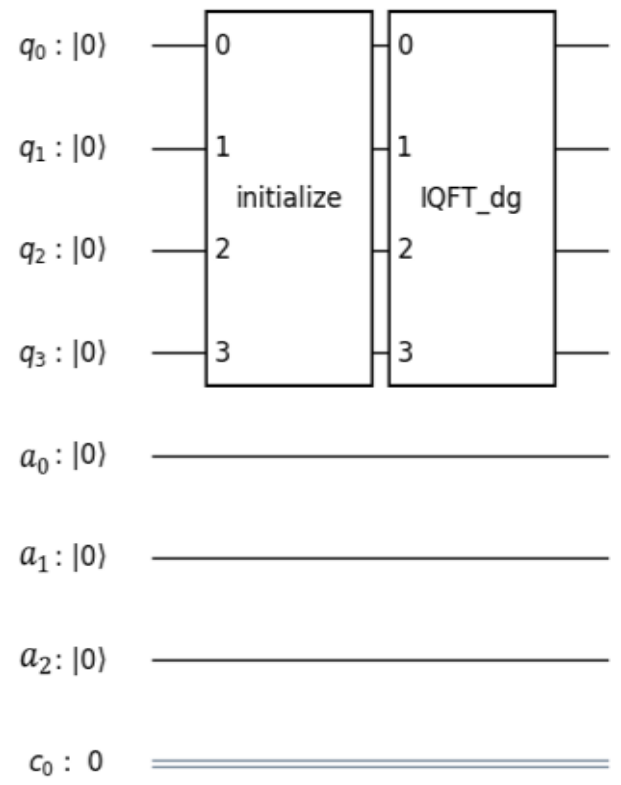

Fig. 3

Notice that so far only the states of first 4 qubits ($q_0$ to $q_3$) have been considered. Later the first ancilla bit $a_0$ will be considered as part of the system since it will mark the unwanted states. The second and the third ancilla bit are not considered as part of the state vector since they are only meant to facilitate CCCCNOT gate (a generalization of Toffoli gate).

Secondly, apply the IQFT circuit for $q_0$ to $q_3$ as shown in Figure 3.

The IQFT subroutine has the following configuration:

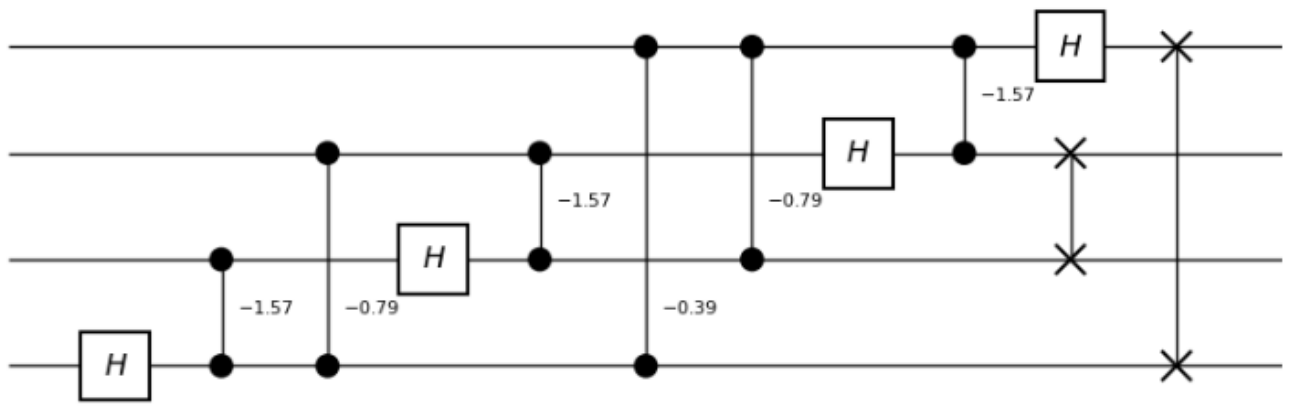

Fig. 4

The output after this stage is:

[ 0.5 +0j, -0.444-0.088j, 0.302+0.125j, -0.132-0.088j, 0+0j, 0.059+0.088j, -0.052-0.125j, 0.018+0.088j, 0+0j, 0.018-0.088j, -0.052+0.125j, 0.059-0.088j, 0+0j, -0.132+0.088j, 0.302-0.125j, -0.444+0.088j ]$^T$

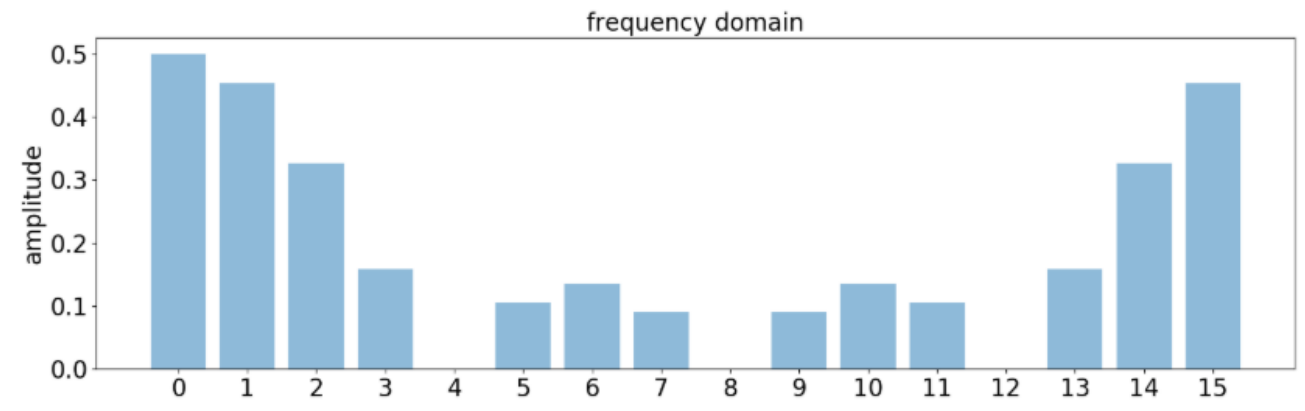

Fig. 5 Absolute Values after IQFT

Notice the bar plot here only shows the absolute values of the complex coefficients and not to be confused with probability amplitude.

### 4.3.1.1 High-Pass Filtering with Qiskit

For the state vector given above, performing high-pass filtering on it is equivalent to eliminate the low frequency components at the first, second, and last row. In other words, the coefficients of those states need to become 0.

The filtering circuit is as shown below:

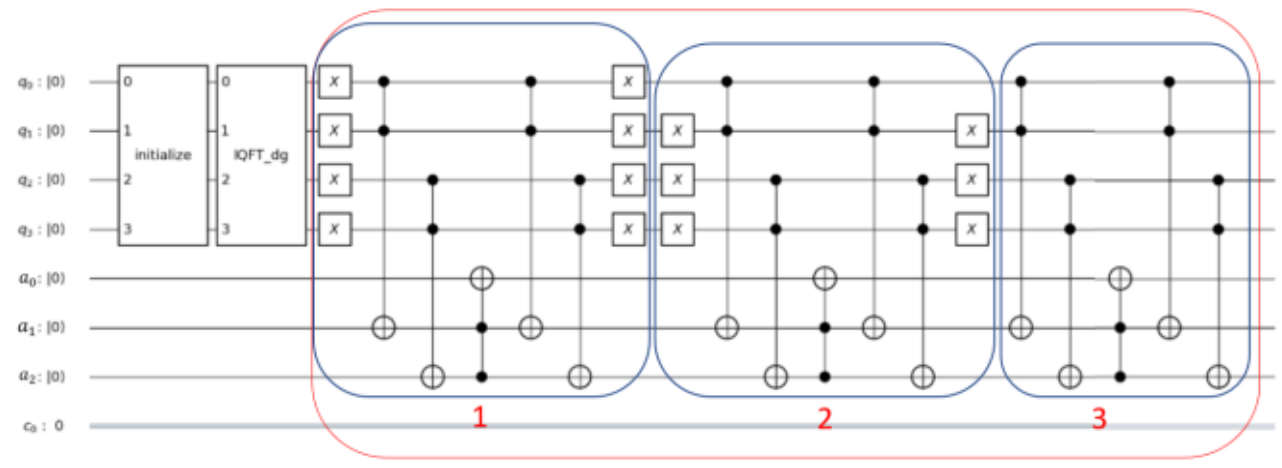

Fig. 6 Filtering circuit

The portion 1, 2, 3 use ancilla bit $a_0$ to mark state $|0000\rangle$, $|0001\rangle$ and $|1111\rangle$ (the first, second and last row of the state vector) with 1 respectively. In other words, when considering $a_0$ as part of the system, the overall state becomes:

[0+0j, 0+0j, 0.302+0.125j, -0.132-0.088j, 0+0j, 0.059+0.088j, -0.052-0.125j, 0.018+0.088j, 0+0j, 0.018-0.088j, -0.052+0.125j, 0.059-0.088j, 0+0j, -0.132+0.088j, 0.302-0.125j, 0+0j, 0.5 -0.j, -0.444-0.088j, 0+0j, 0+0j, 0+0j, 0+0j, 0+0j, 0+0j, 0+0j, 0+0j, 0+0j, 0+0j, 0+0j, 0+0j, 0+0j, -0.444+0.088j ]$^T$

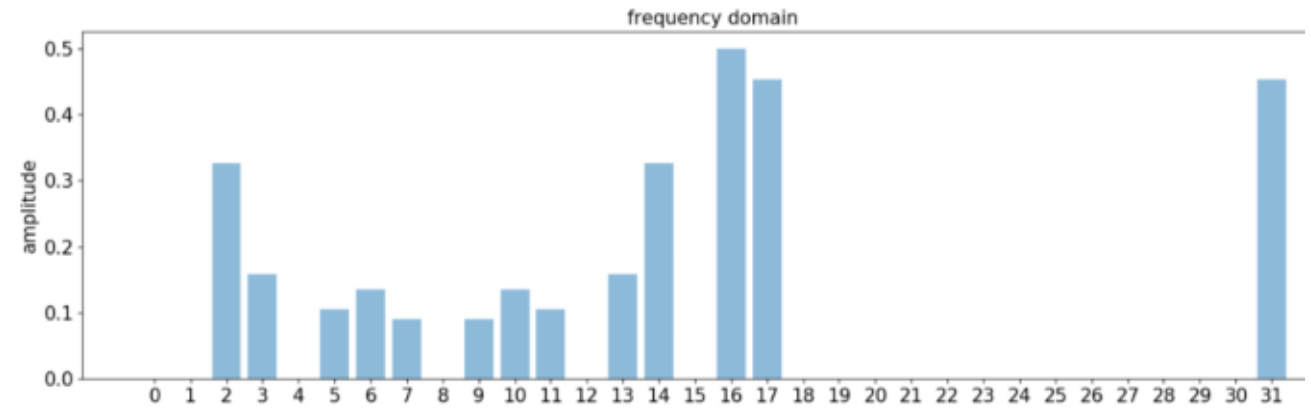

Fig. 7 Absolute Values after Filtering

Notice that the coefficients of $|0000\rangle$, $|0001\rangle$ and $|1111\rangle$ are now the coefficients of $|10000\rangle$ ($|16\rangle$), $|10001\rangle$ ($|17\rangle$) and $|11111\rangle$ ($|31\rangle$).

Now measure $a_0$ in computational basis:

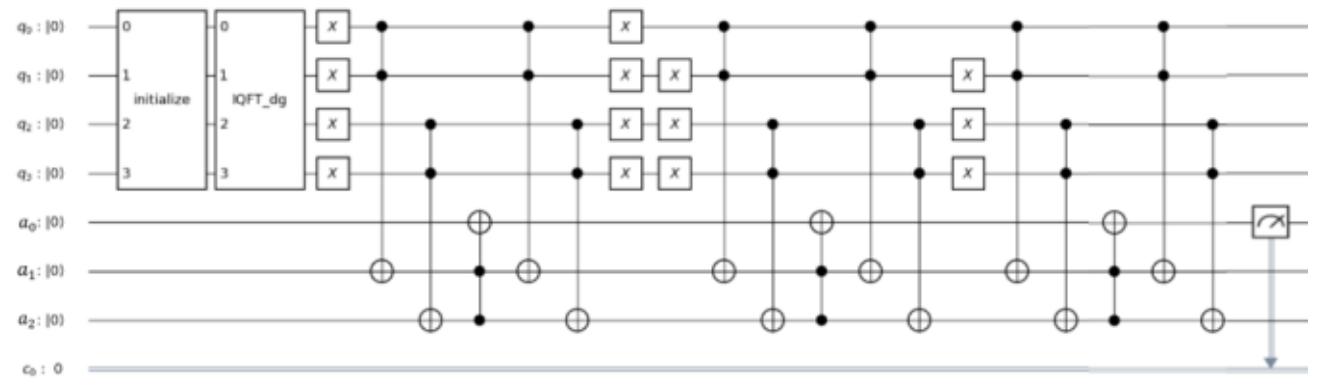

Fig. 8

It will either end up as $|0\rangle$ or $|1\rangle$. When the measurement outcome is $|0\rangle$, the remaining state of the system will become:

[ 0+0j, 0+0j, 0.518+0.215j, -0.227-0.152j, 0+0j, 0.101+0.152j, -0.089-0.215j, 0.03 +0.152j, 0+0j, 0.03-0.152j, -0.089+0.215j, 0.101-0.152j, 0+0j, -0.227+0.152j, 0.518-0.215j, 0+0j, 0+0j, 0+0j, 0+0j, 0+0j, 0+0j, 0+0j, 0+0j, 0+0j, 0+0j, 0+0j, 0+0j, 0+0j, 0+0j, 0+0j, 0+0j, 0+0j]$^T$

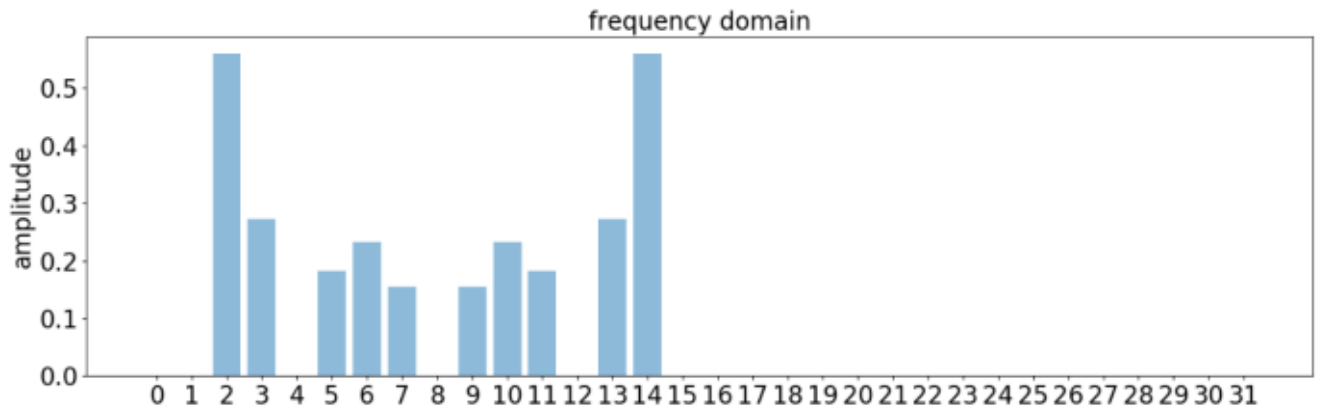

Fig. 9 Absolute Values after Measurement

One can easily compute the probability of measuring $|0\rangle$ which is 0.3392. In fact, the state is obtained after three trials.

Finally, apply QFT:

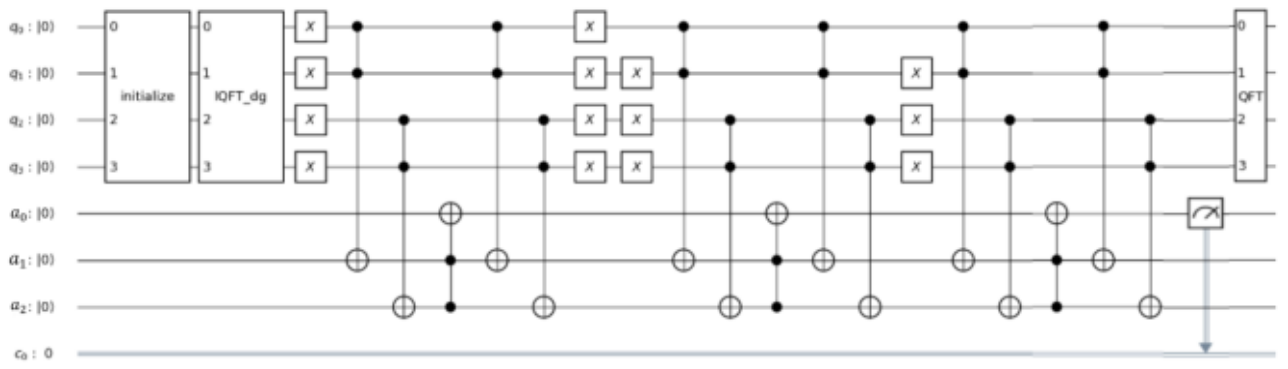

Fig. 10

The QFT subroutine has the following configuration:

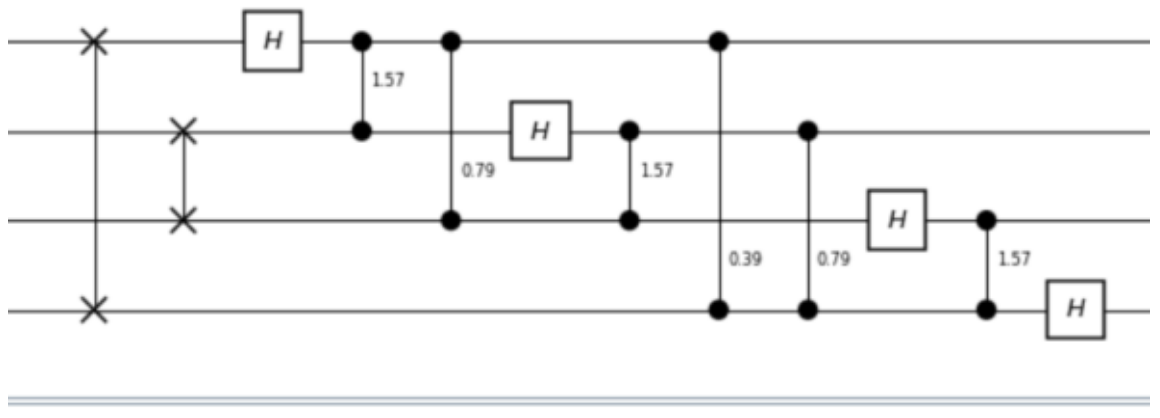

Fig. 11

State vector obtained:

[ 0.167+0j, 0.109+0j, 0.001+0j, -0.139+0j, -0.29 +0j, -0.431+0j, 0.32 +0j, 0.262+0j, 0.262+0j, 0.32+0j, -0.431+0j, -0.29 +0j, -0.139+0j,

| |
|---|
| 0.001+0j, 0.109+0j, 0.167+0j, 0+0j, 0+0j, 0+0j, 0+0j, 0+0j, 0+0j, 0+0j, 0+0j, 0+0j, 0+0j, 0+0j, 0+0j, 0+0j, 0+0j, 0+0j, 0+0j ]$^T$ |

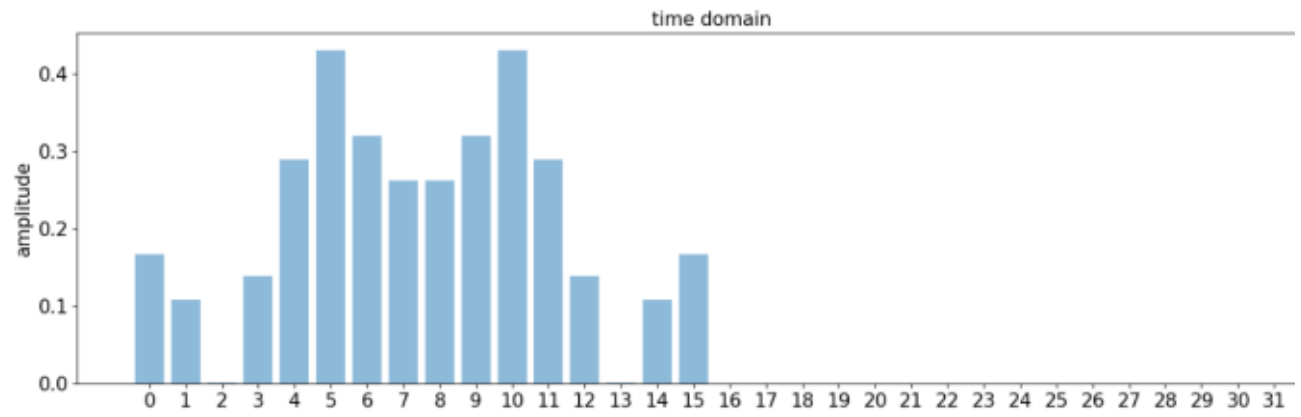

Fig. 12 Absolute Values after QFT

#### 4.3.1.2 Traditional High-Pass Filtering with MATLAB

First, apply a fast Fourier transform (FFT) to the same state vector in MATLAB. The output state vector is:

| |
|---|
| [2.0000 + 0.0000i, -1.7774 - 0.3536i, 1.2071 + 0.5000i, -0.5291 - 0.3536i, 0.0000 + 0.0000i, 0.2362 + 0.3536i, -0.2071 - 0.5000i, 0.0703 + 0.3536i, 0.0000 + 0.0000i, 0.0703 - 0.3536i, -0.2071 + 0.5000i, 0.2362 - 0.3536i, 0.0000 + 0.0000i, -0.5291 + 0.3536i, 1.2071 - 0.5000i, -1.7774 + 0.3536i]$^T$ |

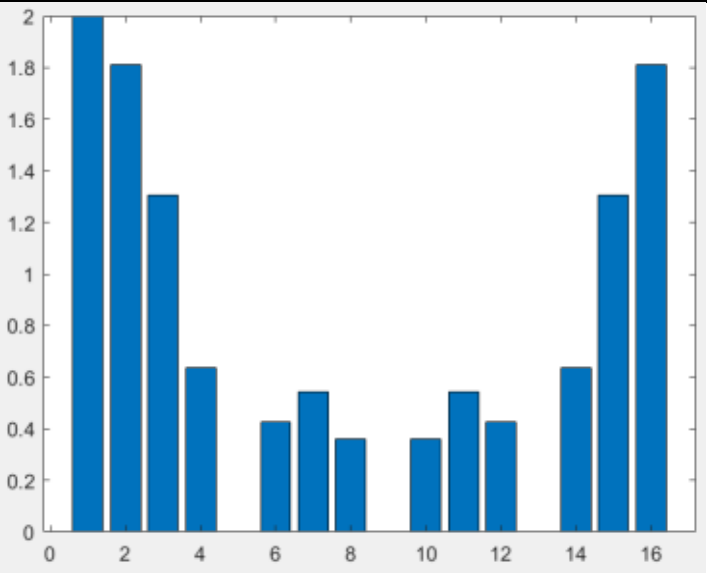

Fig. 13 Absolute Values after FFT

Again, the bar plot only displays the absolute values.

After normalization, the state vector is:

| |
|---|
| [0.5000 + 0.0000i, -0.4444 - 0.0884i, 0.3018 + 0.1250i, -0.1323 - 0.0884i, 0.0000 + 0.0000i, 0.0591 + 0.0884i, -0.0518 - 0.1250i, 0.0176 + 0.0884i, 0.0000 + 0.0000i, 0.0176 - 0.0884i, -0.0518 + 0.1250i, 0.0591 - 0.0884i, 0.0000 + 0.0000i, -0.1323 + 0.0884i, 0.3018 - 0.1250i, -0.4444 + 0.0884i]$^T$ |

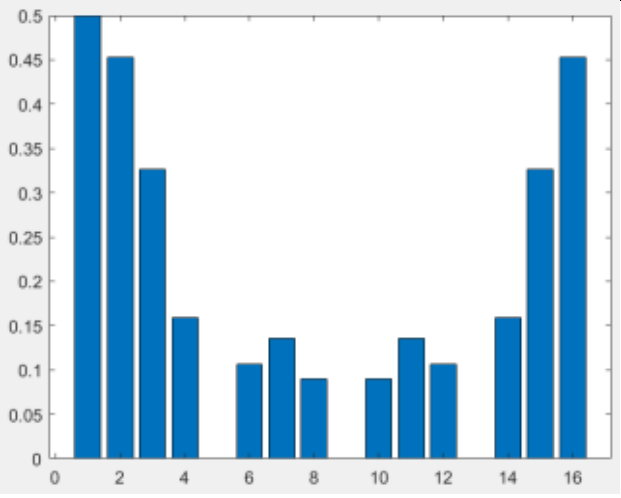

Fig. 14 Absolute Values after FFT (normalized)

Evidently, an IQFT operation is equivalent as an FFT operation after normalization (the minor discrepancy is due to the display rounding error). As a corollary, a QFT operation is equivalent as an inverse fast Fourier transform (IFFT) operation after normalization.

Secondly, mask out the values on the first, second and last row of the unnormalized vector to make them 0:

| |
|---|
| [0.0000 + 0.0000i, 0.0000 + 0.0000i, 0.3018 + 0.1250i, -0.1323 - 0.0884i, 0.0000 + 0.0000i, 0.0591 + 0.0884i, -0.0518 - 0.1250i, 0.0176 + 0.0884i, 0.0000 + 0.0000i, 0.0176 - 0.0884i, -0.0518 + 0.1250i, 0.0591 - 0.0884i, 0.0000 + 0.0000i, -0.1323 + 0.0884i, 0.3018 - 0.1250i, -0.0000 + 0.0000i]$^T$ |

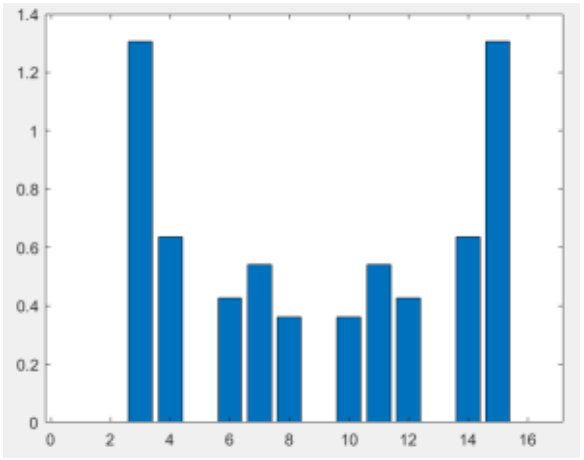

Fig. 15 Absolute Values after Filtering

Now apply inverse fast Fourier transform (IFFT) to the state vector:

| |
|---|
| [0.0972, 0.0634, 0.0009, -0.0808, -0.1692, -0.2509, 0.1866, 0.1528, 0.1528, 0.1866, -0.2509, -0.1692, -0.0808, 0.0009,0.0634, 0.0972]$^T$ |

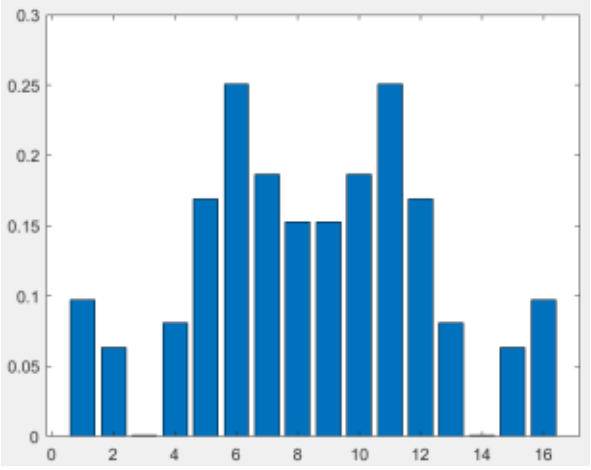

Fig. 16 Absolute Values after IFFT

It can be observed that each entry of the MATLAB final vector is smaller than that of the Qiskit vector by around 1.718 (up to some rounding error). This is due to the renormalization of Qiskit state vector after eliminating the unwanted states' coefficients. The renormalization factor is $\frac{1}{\sqrt{probability\ of\ being\ measured}} = \frac{1}{\sqrt{0.3392}} = 1.717$ which agrees with the numerical result.

It is essentially the same procedure for other types of filtering, except that marked states are changed.

### 4.3.2 Low-Pass Filtering

State $|0110\rangle$, $|0111\rangle$, $|1000\rangle$, $|1001\rangle$ and $|1010\rangle$ are to be marked with the following marking subroutine:

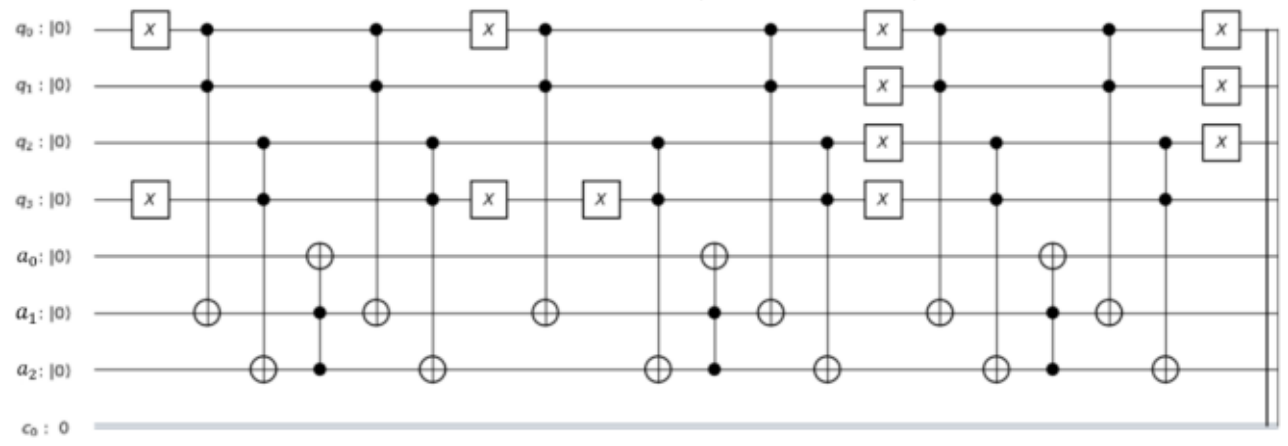

Fig.17a

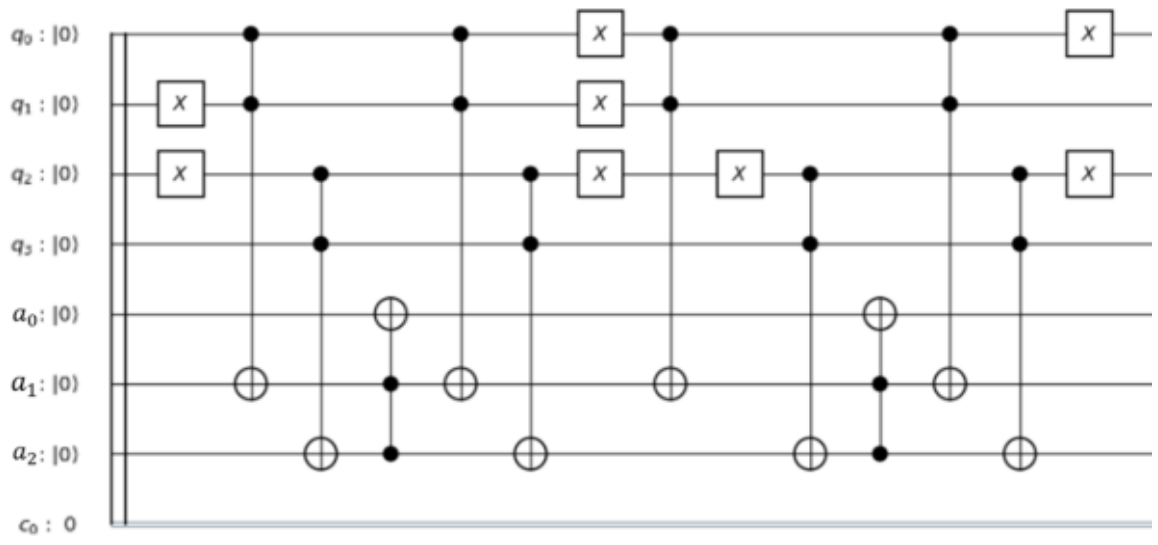

Fig.17b

| | |
|---|---|
| State vector | [0.5-0j, -0.444-0.088j, 0.302+0.125j, -0.132-0.088j, 0+0j, 0.059+0.088j, 0+0j, 0+0j, 0+0j, 0+0j, 0+0j, 0.059-0.088j, 0+0j, -0.132+0.088j, 0.302-0.125j, -0.444+0.088j, 0+0j, 0+0j, 0+0j, 0+0j, 0+0j, 0+0j, -0.052-0.125j, 0.018+0.088j, |
| | 0+0j, 0.018-0.088j, -0.052+0.125j, 0 +0j, 0+0j, 0+0j, 0 +0j, 0+0j]$^T$ |

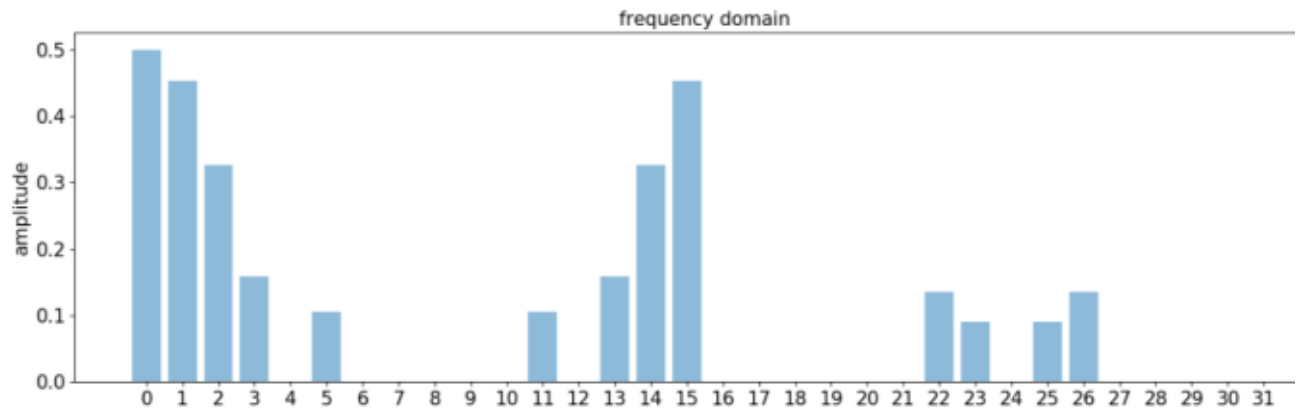

Fig.18 Absolute Values after IQFT

After 1 trial measurement the following state vector is obtained:

| |
|---|
| [ 0.514-0.j, -0.457-0.091j, 0.31 +0.128j, -0.136-0.091j, 0 +0j, 0.061+0.091j, 0+0j, 0+0j, 0+0j, 0 +0j, 0 +0j, 0.061-0.091j, 0 +0j, -0.136+0.091j, 0.31 -0.128j, -0.457+0.091j, 0+0j, 0+0j, 0+0j, 0+0j, 0+0j, 0+0j, 0+0j, 0+0j, 0+0j, 0+0j, 0+0j, 0+0j, 0+0j, 0+0j, 0+0j, 0 +0j]$^T$ |

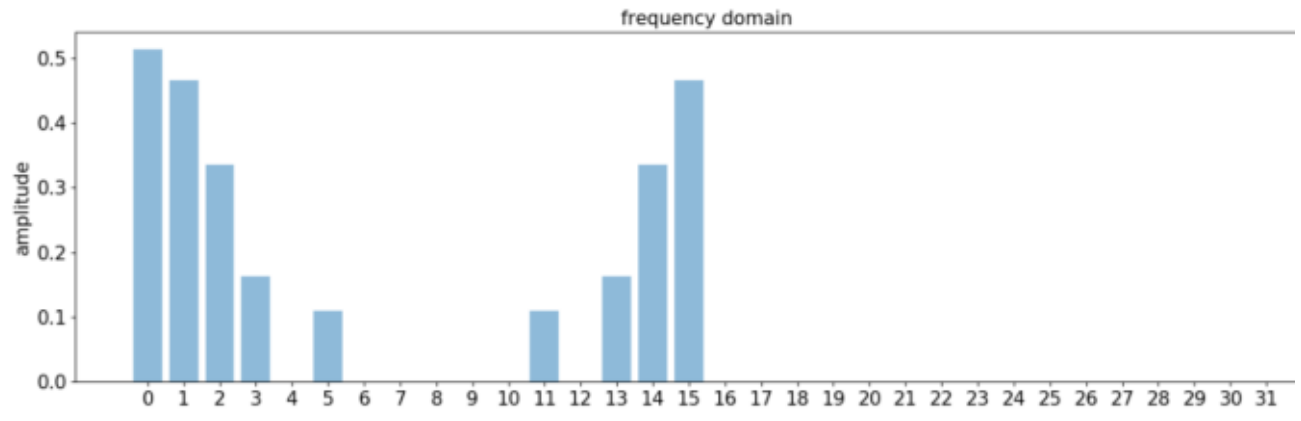

Fig.19 Absolute Values after Measurement

Applying QFT:

| | |
|---|---|
| State vector | [0.018+0j, -0.038+0j, 0.026+0j, 0.019+0j, -0.072+0j, 0.103+0j, 0.424-0j, 0.549-0j, 0.549-0j, 0.424-0j, 0.103+0j, -0.072+0j, 0.019+0j, 0.026+0j, -0.038+0j, 0.018+0j, 0+0j, 0+0j, 0+0j, 0+0j, 0 +0j, 0+0j, 0+0j, 0+0j, 0+0j, 0+0j, 0+0j, 0 +0j, 0 +0j, 0+0j, 0+0j, 0+0j]$^T$ |

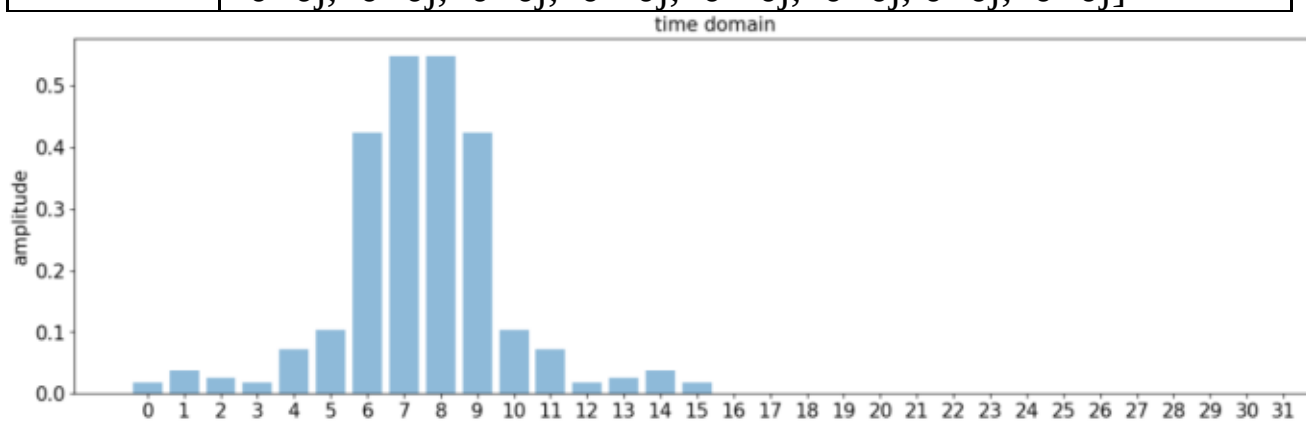

Fig. 20 Absolute Values after QFT

MATLAB simulation and the result of Low-pass filtering and the rest experiments on band-pass and band-stop filtering could be found in the appendix.

### 4.4 Optimized Marking Scheme

Real-world 1-D time series and 2-D images usually correspond to larger state vectors. Therefore, it is not practical to mark individual states one by one. Instead, clustering the unwanted states together then mark those states with much fewer marking subroutines is a much better option.

To illustrate, for the same initial state vector, say now is it desirable to get rid of more low frequency components: $|0000\rangle$, $|0001\rangle$, $|0010\rangle$, $|0011\rangle$, $|1101\rangle$, $|1110\rangle$ and $|1111\rangle$.

Instead of marking 8 states individually, the following could be done:

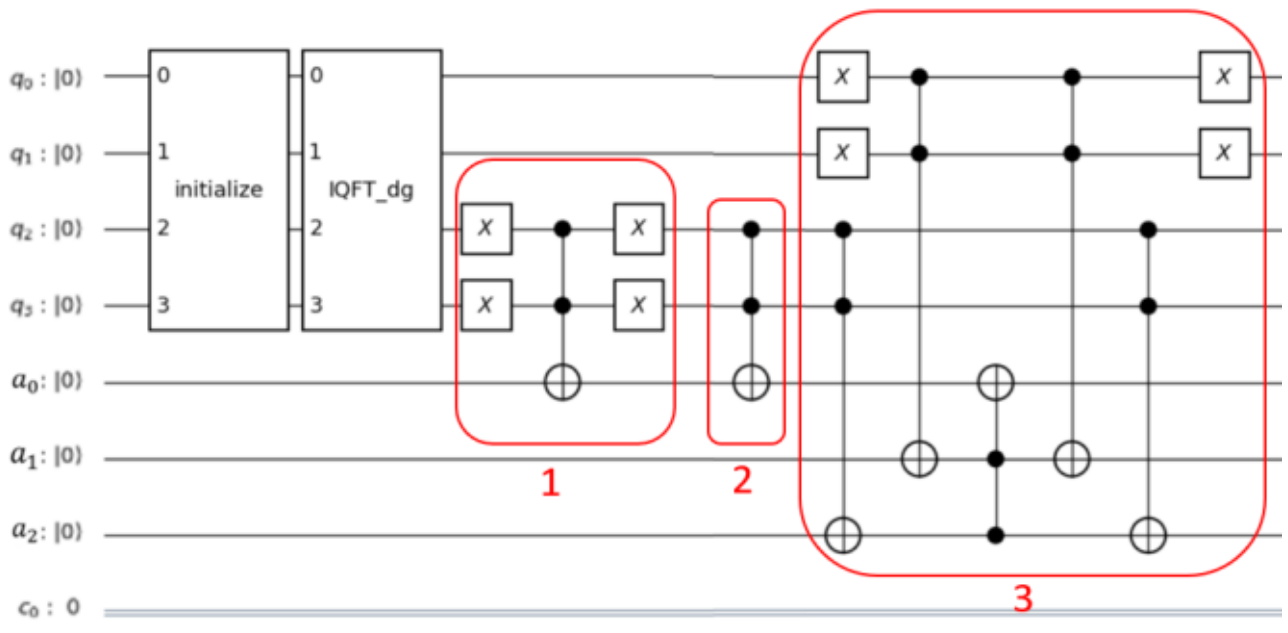

Fig. 21

Subroutine 1 marks state $|0000\rangle$, $|0001\rangle$, $|0010\rangle$, $|0011\rangle$ (denoted as $|00xx\rangle$); Subroutine 2 marks state $|1100\rangle$, $|1101\rangle$, $|1110\rangle$ and $|1111\rangle$ (denoted as $|11xx\rangle$); Subroutine 3 unmarks state $|1100\rangle$.

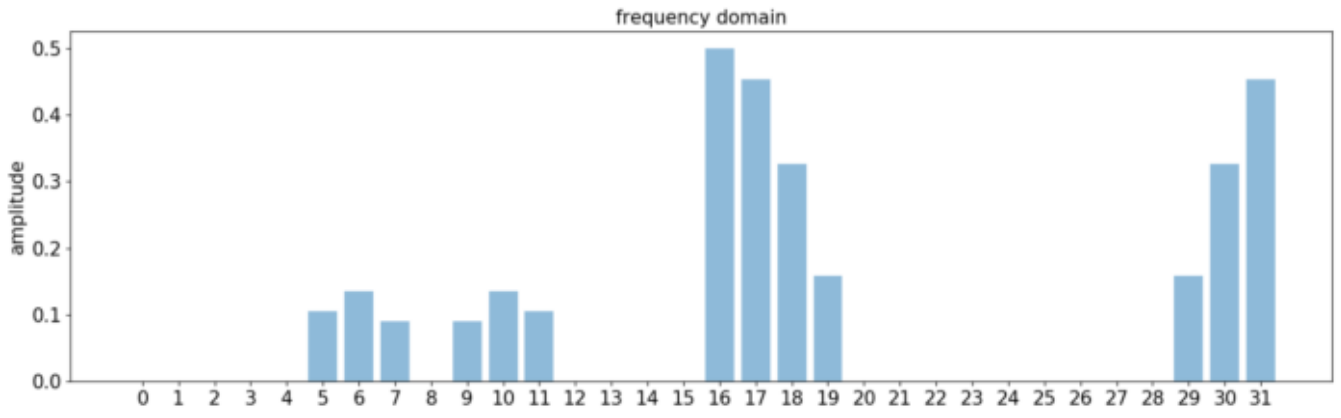

Fig. 22

three subroutines are used to effectively marks 7 states. This technique can be generalized for arbitrarily large state vector. SWAP gate can be helpful for certain specific cluster marking. This could help do filter out arbitrary states' coefficients and achieve other information processing effects.

### 4.5.1 2-D Grey-Scale Image Processing

The method in section 3.2 is used to encode the following 128x128 JPEG grey scale image into the initial state vector.

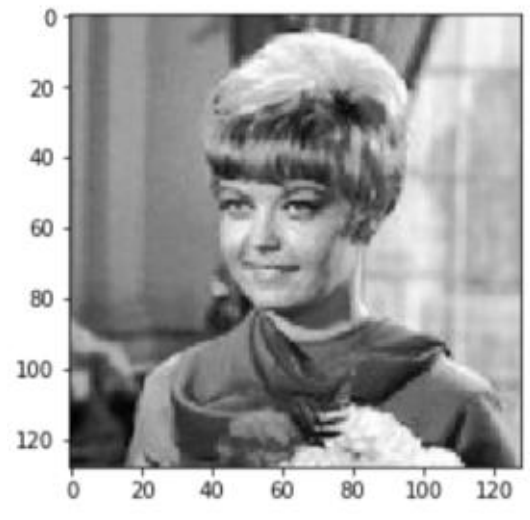

Fig. 23 Original image

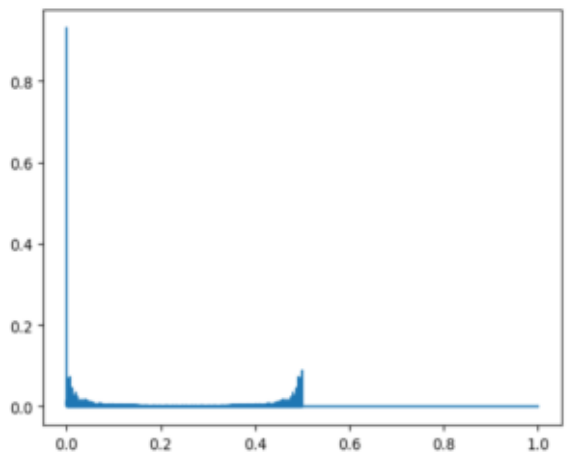

Fig. 24 After QFT

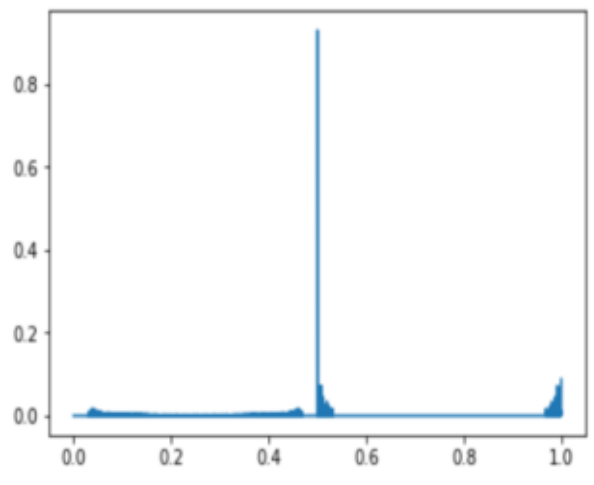

Fig. 25 After marking $|0000xx\ldots xx\rangle$ and $|1111xx\ldots xx\rangle$ states

### 4.5.1.1 Low-Pass Filtering

The low-pass filtering is achieved by marking states $|0000\text{xx} \dots \text{xx}\rangle$ and $|1111\text{xx} \dots \text{xx}\rangle$ where only the first four most significant qubits are marked. This procedure effectively marks $2 \times 2^{10} = 2048$ states at the same time.

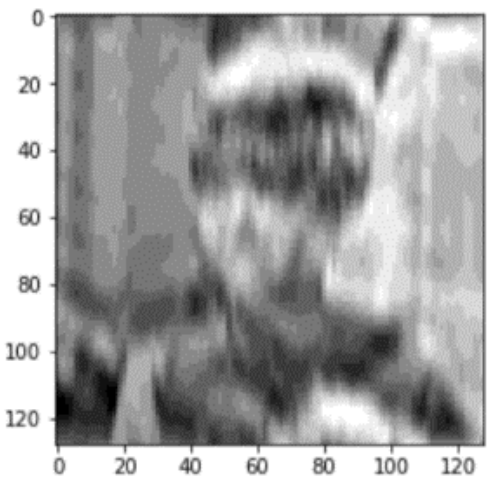

Time taken (overall): 706.39515209198 seconds

Fig. 26

It can be observed that the smoothing effect only occurs long vertical direction. It is because the image matrix is reshaped into 1-D array and IQFT is equivalent to 1-D FFT. However, this is sufficient for certain image processing task.

### 4.5.1.2 High-Pass Filtering

The high-pass filtering is achieved by marking the same states as section 4.5.1.1 does but proceeding with the alternative postselection (measurement) outcome.

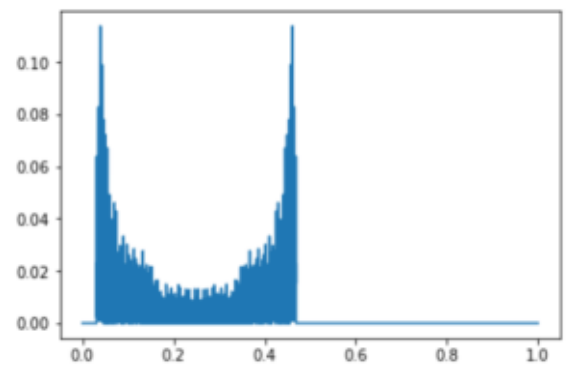

Fig. 27 After measurement

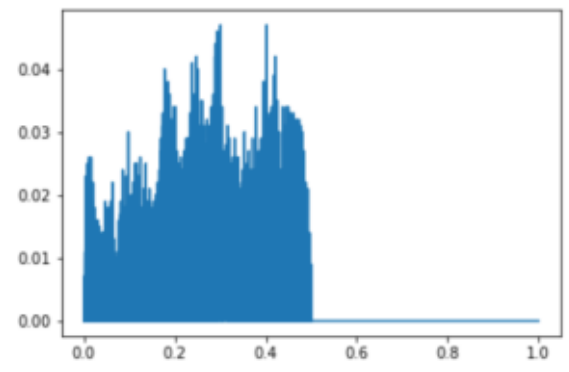

Fig. 28 After QFT

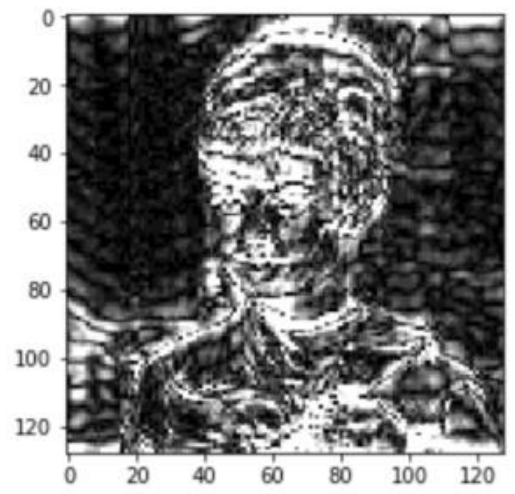

time taken (overall): 693.6424946784973 seconds

Fig. 29

It can be clearly seen that edge detection based on high-pass filtering is implemented successfully.

### 4.5.2 2-D RGB JPEG Image Processing

The following 64x64x3 (12288 pixels) RGB JPEG image is encoded as described in section 3.2:

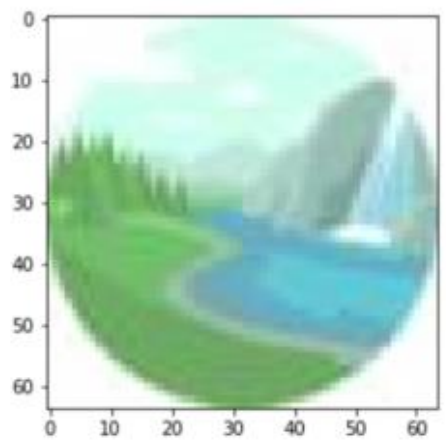

Fig. 30 Original image

The 3-dimensional (3-D) matrix is reshaped into 1-D array and encoded on 14 qubits with last $2^{14} - 12288 = 4096$ states' coefficients left unused. After the processing, the first 12288 states' coefficients are extracted and reshaped into 64x64x3 dimension.

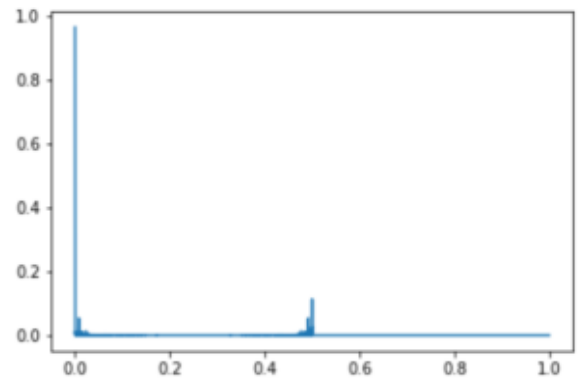

Fig. 31 After QFT

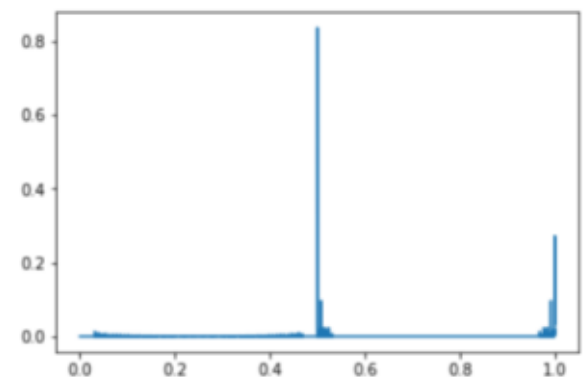

Fig. 32 After marking $|0000\text{xx} \dots \text{xx}\rangle$ and $|1111\text{xx} \dots \text{xx}\rangle$ states

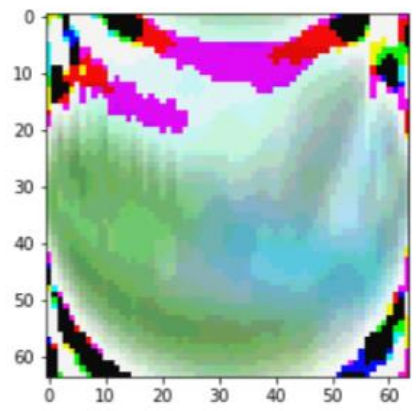

Time taken (overall): 465.44679522514343 seconds

Fig.33

This filtering is achieved by marking states $|0000\text{xx} \dots \text{xx}\rangle$ and $|1111\text{xx} \dots \text{xx}\rangle$ where only the first four most significant qubits are marked. This procedure effectively marks $2 \times 2^{10} = 2048$ states at the same time. Blurring effect can be observed since the high frequency components are masked out. The source of anomalous colored pixels needs further investigation.

## 5 MATRIX TRANSPOSE ALGORITHMS

### 5.1 C-NOT Matrix Transpose

Here a novel method of matrix transpose is introduced. Suppose we have a N by N matrix $A = \begin{bmatrix} a_{11} & \cdots & a_{1N} \\ \vdots & \ddots & \vdots \\ a_{N1} & \cdots & a_{NN} \end{bmatrix}$ where $N^2 = 2^n$ and n is an even positive integer (note that the convention here is different from the previous section). The following quantum circuit is introduced (Figure 34):

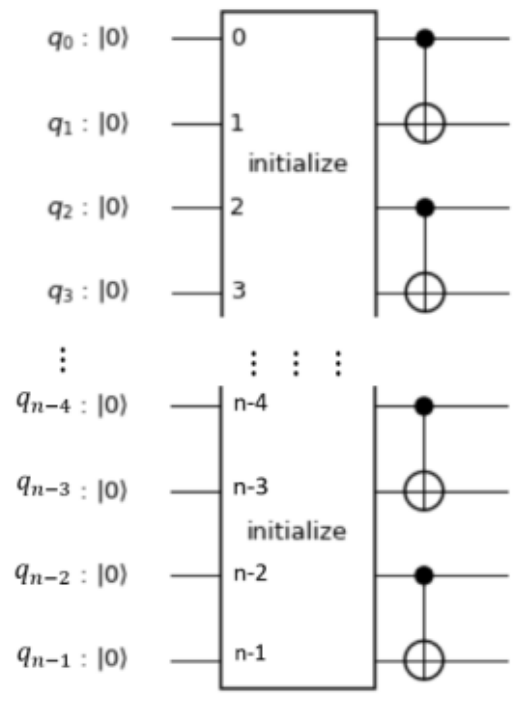

Fig. 34

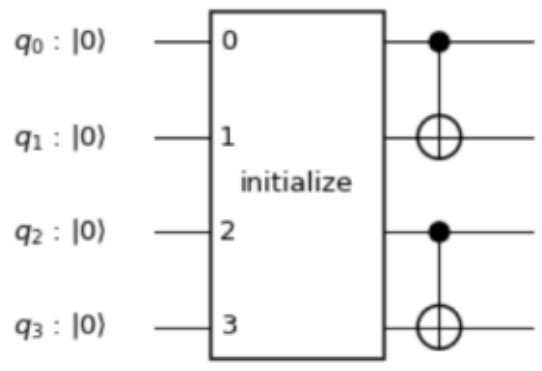

Fig. 35

Notice that this circuit assigns all quantum registers into pairs through controlled-NOT (C-NOT) gates. This gives rise to an effect that when all the control bits are in state $|0\rangle$ no controlled bit will not toggle. This corresponds to the fact that $2^{\frac{n}{2}}$ number of states will not change compared to their states after the initialization stage. As $N^2 = 2^n$, $2^{\frac{n}{2}} = (2^n)^{\frac{1}{2}} = (N^2)^{\frac{1}{2}} = N$. These states are to be encoded with the diagonal elements of the matrix A on their probability amplitude. For all the rest states, it can be simply checked that they are paired up due to the C-NOT gates wiring, in a way that each state will be toggled into the other state within the pair and can never be toggled into any other states.

For illustration, consider a four-qubit matrix-transposing circuit as shown in Figure 35.

Assume the state of the system is

$$\begin{aligned}\Psi = a|0000\rangle + \mathrm{b}|0001\rangle + \mathrm{c}|0010\rangle + \mathrm{d}|0011\rangle + \mathrm{e}|0100\rangle \\ + \mathrm{f}|0101\rangle + \mathrm{g}|0110\rangle + \mathrm{h}|0111\rangle + \mathrm{i}|1000\rangle \\ + \mathrm{j}|1001\rangle + \mathrm{k}|1010\rangle + \mathrm{l}|1011\rangle + \mathrm{m}|1100\rangle \\ + \mathrm{n}|1101\rangle + \mathrm{o}|1110\rangle + \mathrm{p}|1111\rangle\end{aligned}$$

after the initialization stage. Since the register $q_0$ and $q_2$ are the control bits, the states that do not change after the C-NOT gates are $|0000\rangle$, $|0010\rangle$, $|1000\rangle$ and $|1010\rangle$ where the control bits are both zero (the least significant bit refers to the quantum register at the top and the most significant bit refers to the quantum register at the bottom). It is simple to check that the following pairs of states will be toggled from one to the other:
$|0001\rangle$ and $|0011\rangle$; $|1001\rangle$ and $|1011\rangle$;
$|0100\rangle$ and $|1100\rangle$; $|0110\rangle$ and $|1110\rangle$;
$|0101\rangle$ and $|1111\rangle$; $|0111\rangle$ and $|1101\rangle$.

Arranging the states:

$$\begin{aligned}\Psi' = a|0000\rangle + \mathrm{d}|0001\rangle + \mathrm{c}|0010\rangle + \mathrm{b}|0011\rangle + \mathrm{m}|0100\rangle \\ + \mathrm{p}|0101\rangle + \mathrm{o}|0110\rangle + \mathrm{n}|0111\rangle + \mathrm{i}|1000\rangle \\ + \mathrm{l}|1001\rangle + \mathrm{k}|1010\rangle + \mathrm{j}|1011\rangle + \mathrm{e}|1100\rangle \\ + \mathrm{h}|1101\rangle + \mathrm{g}|1110\rangle + \mathrm{f}|1111\rangle\end{aligned}$$

This is the states at the end of the circuit.

Assign all the basis states into a 4-by-4 matrix B in such a way that the invariant states are on the diagonal entries and the toggled states are on the two sides of the diagonal and the pairs are at the position symmetrical about the diagonal:

$$B = \begin{bmatrix} |0000\rangle & |0001\rangle & |1001\rangle & |0100\rangle \\ |0011\rangle & |0010\rangle & |0110\rangle & |0101\rangle \\ |1011\rangle & |1110\rangle & |1000\rangle & |1101\rangle \\ |1100\rangle & |1111\rangle & |0111\rangle & |1010\rangle \end{bmatrix}$$

Notice that there are multiple ways of constructing such a matrix.

Putting the corresponding coefficients (after initialization stage) into another 4-by-4 matrix C:

$$C = \begin{bmatrix} a & b & j & e \\ d & c & g & f \\ l & o & i & n \\ m & p & h & k \end{bmatrix}$$

At the end of the circuit,

$$B' = \begin{bmatrix} |0000\rangle & |0011\rangle & |1011\rangle & |1100\rangle \\ |0001\rangle & |0010\rangle & |1110\rangle & |1111\rangle \\ |1001\rangle & |0110\rangle & |1000\rangle & |0111\rangle \\ |0100\rangle & |0101\rangle & |1101\rangle & |1010\rangle \end{bmatrix}$$

It can be observed that B' is the matrix transpose of B. If measurement is carried out, it can be observed that the probability amplitudes of the paired basis states are interchanged. In other words, if other operations are done on this circuit it is equivalent to do the operations to the transposed information:

$$C = \begin{bmatrix} a & b & j & e \\ d & c & g & f \\ l & o & i & n \\ m & p & h & k \end{bmatrix}$$

### 5.1.1 Simulation on Qiskit

The four-qubit case simulation is run on IBM Qiskit. To begin with, prepare the initial state of the system with an initialization unitary circuit to encode

$$X = \begin{bmatrix} 0 & 1 & 2 & 3 \\ 4 & 5 & 6 & 7 \\ 8 & 9 & 10 & 11 \\ 12 & 13 & 14 & 15 \end{bmatrix}$$

First normalize X to be

$$X' = \begin{bmatrix} 0 & 0.091 & 0.129 & 0.158 \\ 0.183 & 0.204 & 0.224 & 0.242 \\ 0.258 & 0.274 & 0.289 & 0.303 \\ 0.316 & 0.329 & 0.342 & 0.354 \end{bmatrix}$$

with a normalization factor of $\frac{1}{120}$. Then encode X' onto B by assigning each value onto the basis state at the same position.

The resultant state is

$$\begin{aligned}\Psi' = 0|0000\rangle + 0.091|0001\rangle + 0.204|0010\rangle \\ + 0.183|0011\rangle + 0.158|0100\rangle \\ + 0.242|0101\rangle + 0.224|0110\rangle \\ + 0.342|0111\rangle + 0.289|1000\rangle \\ + 0.129|1001\rangle + 0.354|1010\rangle \\ + 0.258|1011\rangle + 0.316|1100\rangle \\ + 0.303|1101\rangle + 0.274|1110\rangle \\ + 0.329|1111\rangle\end{aligned}$$

As proven before, this is always possible to implement such a unitary circuit:

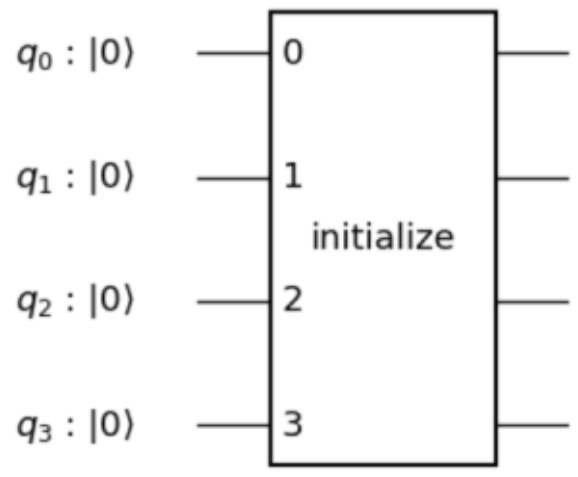

Fig. 36

The state vector after the initialization stage is:

| |
|---|
| [0, 0.091, 0.204, 0.183, 0.158, 0.242, 0.224, 0.342, 0.289, 0.129, 0.354, 0.258, 0.316, 0.303, 0.274, 0.329]$^T$ |

which is the same as $\Psi'$.

Now input the state vector into the C-NOT gate component as figure 35 shows. The final output state vector is:

[0, 0.183, 0.204, 0.091, 0.316, 0.329, 0.274, 0.303, 0.289, 0.258, 0.354, 0.129, 0.158, 0.342, 0.224, 0.242]$^T$

The corresponding probability amplitudes are:

[0, 0.033489, 0.041616, 0.008281, 0.099856, 0.108241, 0.075076, 0.091809, 0.083521, 0.066564, 0.125316, 0.016641, 0.024964, 0.116964, 0.050176, 0.058564]

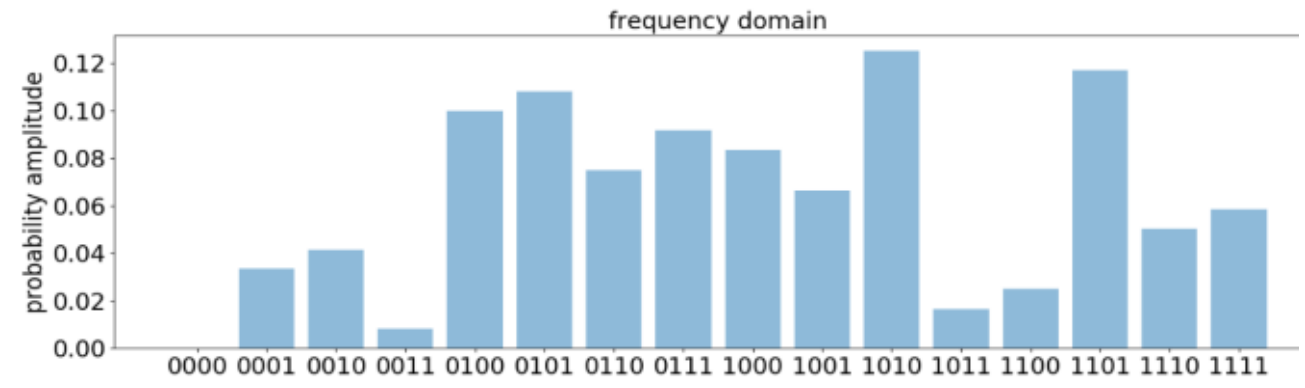

Fig. 37

It can be checked that if we still treat the final output as matrix B, the measured matrix will be:

$$\begin{bmatrix} 0 & 0.033489 & 0.066564 & 0.099856 \\ 0.008281 & 0.041616 & 0.075076 & 0.108241 \\ 0.016641 & 0.050176 & 0.083521 & 0.116964 \\ 0.024964 & 0.058564 & 0.091809 & 0.125316 \end{bmatrix}$$

Since this result is already after squaring the coefficients of the basis states, to recover the denormalized matrix, multiply each entry by 120:

$$\begin{bmatrix} 0 & 4.01868 & 7.98768 & 11.98272 \\ 0.99372 & 4.99392 & 9.00912 & 12.98892 \\ 1.99692 & 6.02112 & 10.02252 & 14.03568 \\ 2.99568 & 7.02768 & 11.01708 & 15.03792 \end{bmatrix}$$

The small discrepancy is due to the rounding-off error which will not occur in the actual physical system.

For implementation, prepare the circuit as per described above but use a switch to by-pass all the C-NOT gates. When matrix transpose is required, toggle the switch to put the C-NOT gates inside the circuit:

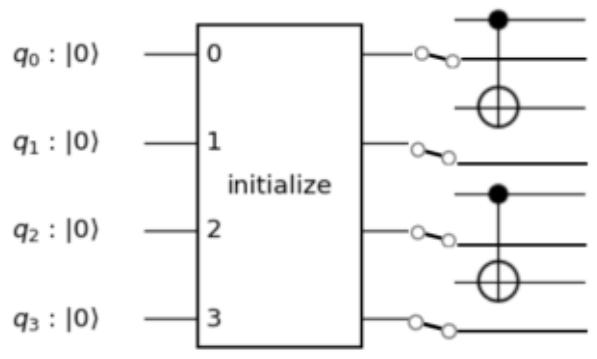

Fig. 38

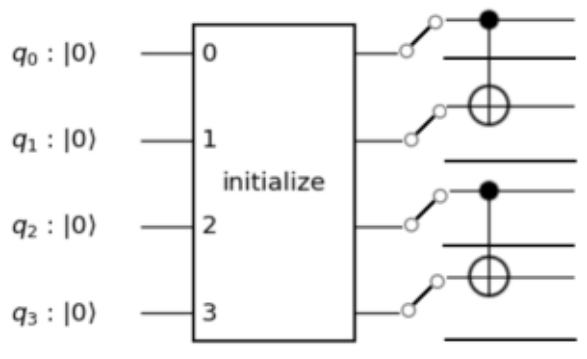

Fig. 39

This scheme involves only multiple bipartite entanglements due to C-NOT gates. This means it is experimentally much easier to implement and maintain than other scheme which might involve multipartite entanglements.

## 5.2 SWAP Matrix Transpose

SWAP Matrix transpose has the following circuit:

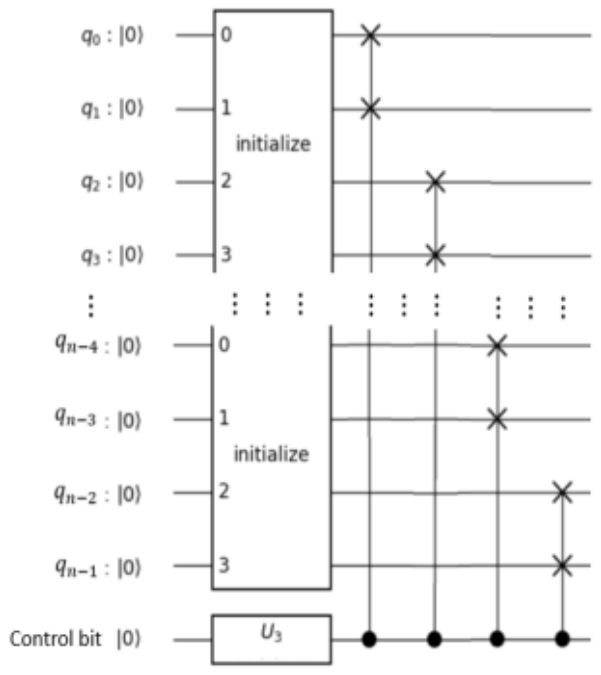

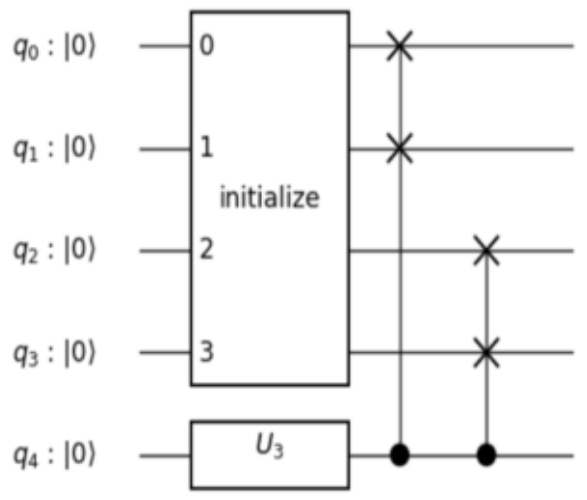

Fig.40 general layout Fig.41 4-qubit layout

$$D = \begin{bmatrix} |0000\rangle & |0001\rangle & |0100\rangle & |0110\rangle \\ |0010\rangle & |0011\rangle & |0101\rangle & |0111\rangle \\ |1000\rangle & |1010\rangle & |1100\rangle & |1101\rangle \\ |1001\rangle & |1011\rangle & |1110\rangle & |1111\rangle \end{bmatrix}$$

There are multiple ways of constructing such a matrix as well.

Say we still want to transpose

$$X = \begin{bmatrix} 0 & 1 & 2 & 3 \\ 4 & 5 & 6 & 7 \\ 8 & 9 & 10 & 11 \\ 12 & 13 & 14 & 15 \end{bmatrix}$$

Now initialize our state vector it as:

$$[0, \frac{1}{\sqrt{1240}}, \frac{4}{\sqrt{1240}}, \frac{5}{\sqrt{1240}}, \frac{2}{\sqrt{1240}}, \frac{6}{\sqrt{1240}}, \frac{3}{\sqrt{1240}}, \frac{7}{\sqrt{1240}},$$
$$\frac{8}{\sqrt{1240}}, \frac{12}{\sqrt{1240}}, \frac{9}{\sqrt{1240}}, \frac{13}{\sqrt{1240}}, \frac{10}{\sqrt{1240}}, \frac{11}{\sqrt{1240}}, \frac{14}{\sqrt{1240}}, \frac{15}{\sqrt{1240}}]$$

When the input parameters of $U_3$ gate are 0, 0, 0, $U_3$ gate is an identity gate. The state of control bit $q_4$ remains $|0\rangle$, the SWAP gate is disenabled and hence we have our initial state vector. When the input parameters of $U_3$ gate are $\pi, 0, \pi$, $U_3$ gate is a X-gate. The state of control bit $q_4$ is now $|1\rangle$ , the SWAP gate is enabled, and our state vector becomes:

[0, 0.114, 0.028, 0.142, 0.227, 0.256, 0.341, 0.369, 0.057, 0.085, 0.17, 0.199, 0.284, 0.398, 0.312, 0.426]$^T$

Multiplying by the scaling factor:

[0, 4.01435424, 0.98598174, 5.00033599, 7.99349486, 9.01469023, 12.0078491, 12.99383084, 2.00717712, 2.99315887, 5.98631773, 7.00751311, 10.00067198, 14.01502622, 10.98665372, 15.00100797]$^T$

Put it back in matrix form:

$$\begin{bmatrix} 0 & 4.01435424 & 7.99349486 & 12.0078491 \\ 0.98598174 & 5.00033599 & 9.01469023 & 12.99383084 \\ 2.00717712 & 5.98631773 & 10.00067198 & 14.01502622 \\ 2.99315887 & 7.00751311 & 10.98665372 & 15.00100797 \end{bmatrix}$$

which is the result we want.

## 5.3 General Matrix Transpose

So far, only the transpose for $2^{\frac{n}{2}} \times 2^{\frac{n}{2}}$ matrices has been demonstrated. For a more general matrix, what needs to be done is to construct a $2^{\frac{n}{2}} \times 2^{\frac{n}{2}}$ matrix which can contain the entire matrix of interest with minimum $n$ possible. For example, for

$E = \begin{bmatrix} a & b & c \\ d & e & f \end{bmatrix}$, it needs to be put into

$$E' = \begin{bmatrix} a & b & c & 0 \\ d & e & f & 0 \\ 0 & 0 & 0 & 0 \\ 0 & 0 & 0 & 0 \end{bmatrix}$$

and do transpose.

### 5.4 Potential Application

Yao et al. have proposed a quantum edge detection algorithm in their paper Quantum Image Processing and Its Application to Edge Detection: Theory and Experiment [11]. While the circuit used is simplistic, it requires to prepare the initial state of the system twice for horizontal scan and vertical scan respectively. The effects of the proposed matrix transpose subroutines are equivalent to a vertical-scan state preparation.

The advantage of doing so is that the overhead of encoding information onto qubits is drastically decreased. It is well-known that in general the large overhead of quantum algorithm severely compromises the exponential speedup provided by quantum algorithms and hence defeat their purpose. Therefore, the matrix transpose circuit truly helps the edge detection algorithm to shine in its full glory.

Furthermore, as mentioned the current proposed scheme for 2-D image filtering has its drawback since it treats the images as 1-D array. With the matrix transpose subroutine, it is possible to achieve the same effect as a 2-D FFT (2DFFT) operation since $2\mathrm{DFFT} = \mathrm{FFT}((\mathrm{FFT}(\mathrm{v}))^T)^T$ where v is the reshaped 1-D array from a 2-D image matrix.

However, the proposed edge detection algorithm works by subtracting the adjacent states to find the differences the adjacent pixel values. This means on top of the encoding rules given above, the states in the same rows should be in consecutive order. Clearly it is not possible in four-qubit case. However, as the number of qubits gets larger, there will be more flexibility in terms of satisfying the encoding rules as in both scheme each 2-qubit gate (C-NOT and SWAP gate) can be connected to any two of the qubits to form a pair. Therefore, it might be possible for this condition to be satisfied simultaneously with the rules. This needs further investigation to be proved.

## 6 CONCLUSION

As shown, the proposed schemes are able to process 1-D time series and 2-D images correctly and effectively. However, since the experiments were done by simulation, the speedup could not be seen as promised if the algorithm is implemented on a real quantum computer. Moreover, this simulation provides noise-free condition for the proposed schemes to work. If the schemes are carried out on a real quantum computer, the environmental noise will reduce the fidelity of the processed data. As the size of data gets larger and larger, the algorithm becomes more and more demanding to low noise condition. Therefore, it is imperative to develop counter measures such that quantum algorithms can be more robust and tolerant against non-ideal working environment.

## ACKNOWLEDGMENT

The author is grateful for two images used in section 4.5.1 and 4.5.2 which are adapted from http://www.fmwconcepts.com/imagemagick/levelcolors/index.php and https://www.shareicon.net/data/128x128/2016/09/16/829667_nature_512x512.png (preprocessed).

The author is grateful for the advice given by user DaftWullie on StackExchange.

The simulations in this paper are implemented by using IBM Qiskit [12] and MATLAB. The quantum circuit diagrams are constrcuted from raw cirucit diagrams from IBM Qiskit.

We wish to acknowledge the funding support for this project from Nanyang Technological University under the Undergraduate Research Experience on CAmpus (URECA) programme.

## APPENDIX

All simulations have the same initial state as in section 4.2. All Qiskit simulations have the same state vector after IQFT as in section 4.3.1.1 and all MATLAB simulations have the same state vector after FFT as in section 4.3.1.2.

### MATLAB Low-Pass Filtering

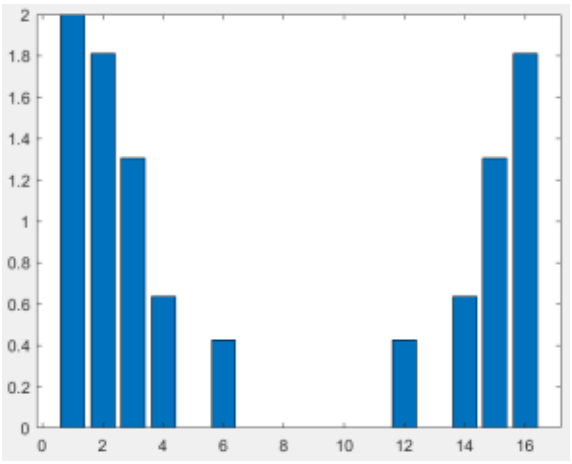

Fig. 42 Absolute Values after Filtering

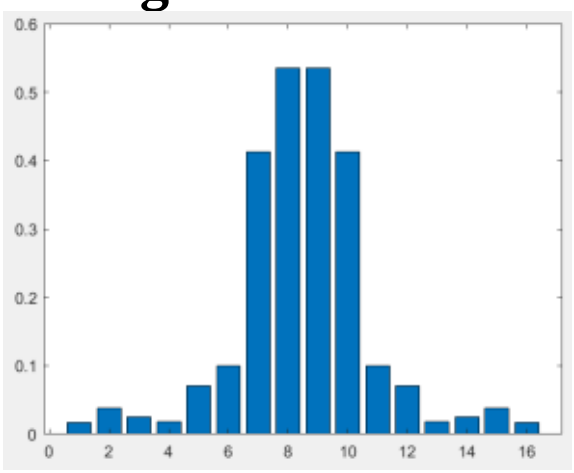

Fig. 43 Absolute Values after IFFT

| | |
|---|---|
| State vector after filtering | [2.0000 + 0.0000i, -1.7774 - 0.3536i, 1.2071 + 0.5000i, -0.5291 - 0.3536i, 0.0000 + 0.0000i, 0.2362 + 0.3536i, 0.0000 + 0.0000i, 0.0000 + 0.0000i, 0.0000 + 0.0000i, 0.0000 + 0.0000i, 0.0000 + 0.0000i, 0.2362 - 0.3536i, 0.0000 + 0.0000i, -0.5291 + 0.3536i, 1.2071 - 0.5000i, -1.7774 + 0.3536i]$^T$ |
| State vector after IFFT | [0.0171, -0.0375, 0.0250, 0.0183, -0.0701, 0.1000, 0.4125, 0.5347, 0.5347, 0.4125, 0.1000, -0.0701, 0.0183, 0.0250, -0.0375, 0.0171]$^T$ |

### Qiskit Band-Pass Filtering

| | |
|---|---|
| State vector after filtering | [0+0j, 0+0j, 0.302+0.125j, -0.132-0.088j, 0+0j, 0.059+0.088j, -0.052-0.125j, 0+0j, 0+0j, 0+0j, -0.052+0.125j, 0.059-0.088j, 0+0j, -0.132+0.088j, 0.302-0.125j, 0+0j, 0.5-0j, -0.444-0.088j, 0+0j, 0+0j, 0+0j, 0+0j, 0.018+0.088j, 0+0j, 0.018-0.088j, 0+0j, 0+0j, 0+0j, 0+0j, 0+0j, 0+0j, -0.444+0.088j]$^T$ |

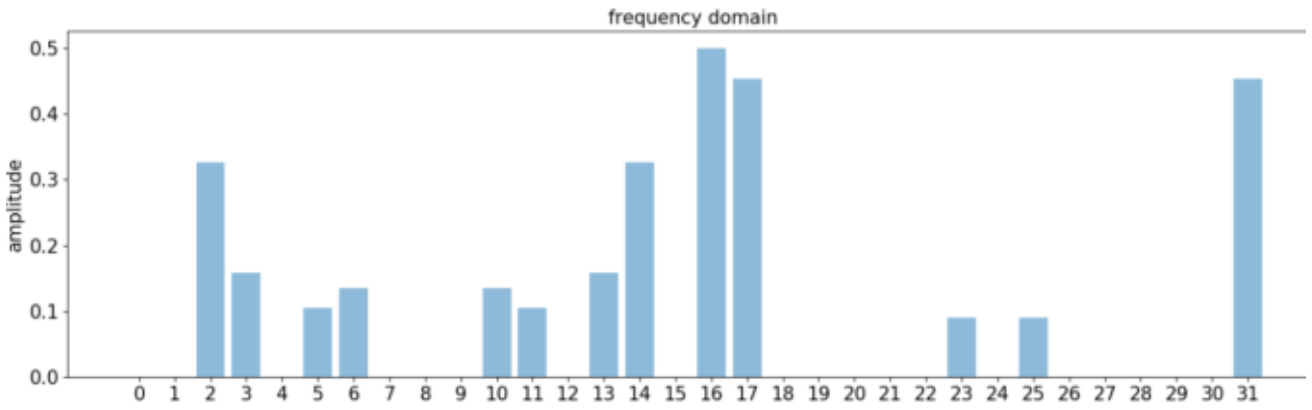

Fig. 44 Absolute Values after Filtering

| | |
|---|---|
| State vector After measurement (1 trial) | [0+0j, 0+0j, 0.531+0.22j, -0.233-0.155j, 0+0j, 0.104+0.155j, -0.091-0.22j, 0+0j, 0+0j, 0+0j, -0.091+0.22j, 0.104-0.155j, 0+0j, -0.233+0.155j, 0.531-0.22j, 0+0j, 0+0j, 0+0j, 0+0j, 0+0j, 0+0j, 0+0j, 0+0j, 0+0j, 0+0j, 0+0j, 0+0j, 0+0j, 0+0j, 0+0j, 0+0j, 0+0j]$^T$ |

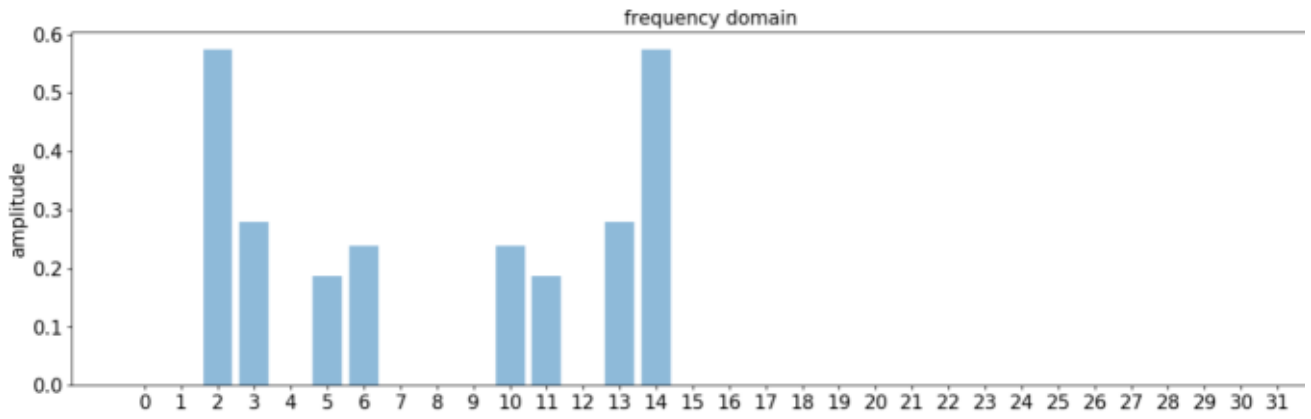

Fig. 45 Absolute Values after Measurement (1trail)

| | |
|---|---|
| State vector after QFT | [0.155+0.j, 0.155+0.j, -0.064+0.j, -0.064+0.j, -0.375+0.j, -0.375+0.j, 0.284+0.j, 0.284+0.j, 0.284+0.j, 0.284+0.j, -0.375+0.j, -0.375+0.j, -0.064+0.j, -0.064+0.j, 0.155+0j, 0.155+0j, 0+0j, 0+0j, 0+0j, 0+0j, 0+0j, 0+0j, 0+0j, 0+0j, 0+0j, 0+0j, 0+0j, 0+0j, 0+0j, 0+0j, 0+0j, 0+0j ]$^T$ |

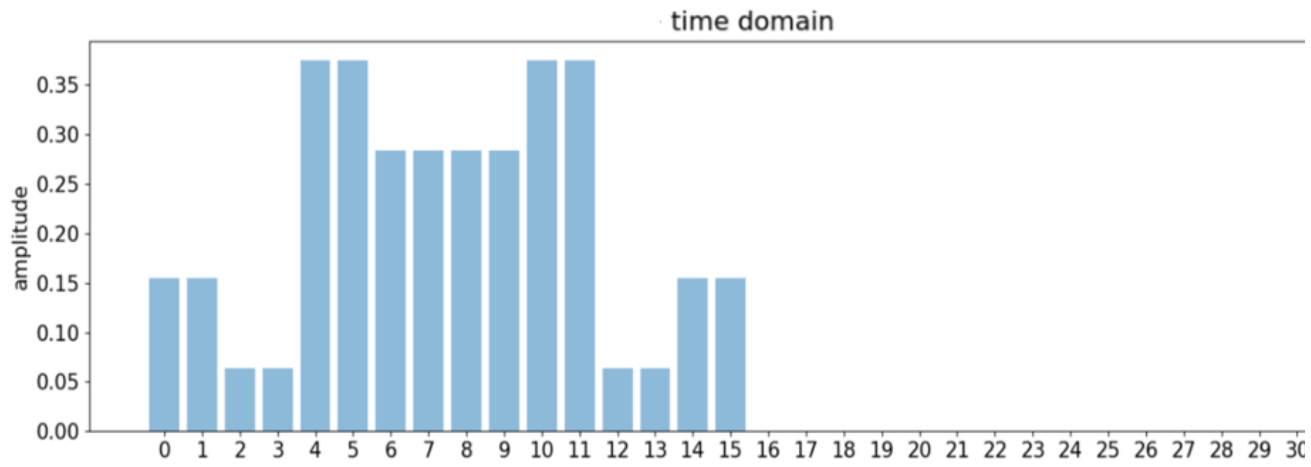

Fig. 46 Absolute Values after QFT

## MATLAB Band-Pass Filtering

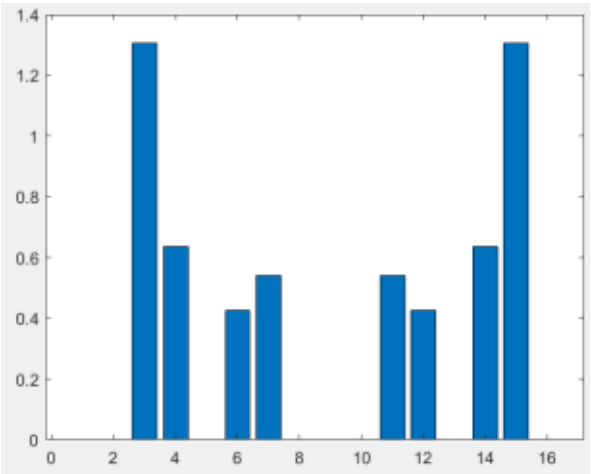

Fig. 47 Absolute Values after Filtering

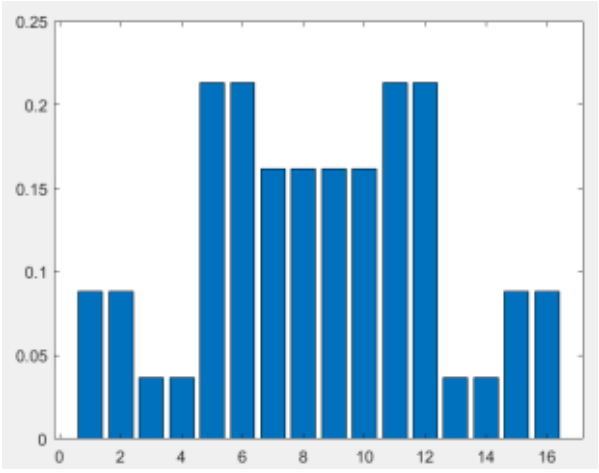

Fig. 48 Absolute Values after IFFT

| | |
|---|---|
| State vector after filtering | [0.0000 + 0.0000i, 0.0000 + 0.0000i, 1.2071 + 0.5000i, -0.5291 - 0.3536i, 0.0000 + 0.0000i, 0.2362 + 0.3536i, -0.2071 - 0.5000i, 0.0000 + 0.0000i, 0.0000 + 0.0000i, 0.0000 + 0.0000i, -0.2071 + 0.5000i, 0.2362 - 0.3536i, 0.0000 + 0.0000i, -0.5291 + 0.3536i, 1.2071 - 0.5000i, 0.0000 + 0.0000i]$^T$ |
| State vector | [0.0884, 0.0884, -0.0366, -0.0366, -0.2134, -0.2134, 0.1616, 0.1616, 0.1616, 0.1616, -0.2134, -0.2134, -0.0366, -0.0366, 0.0884, 0.0884]$^T$ |
| after IFFT | |

## Qiskit Band-Stop (Notch) Filtering

| | |
|---|---|
| State vector after filtering | [0.5+0.j, -0.444-0.088j, 0+0j, 0+0j, 0+0j, 0.059+0.088j, -0.052-0.125j, 0.018+0.088j, 0+0j, 0.018-0.088j, -0.052+0.125j, 0.059-0.088j, 0+0j, 0+0j, 0+0j, -0.444+0.088j, 0+0j, 0+0j, 0.302+0.125j, -0.132-0.088j, 0+0j, 0+0j, 0+0j, 0+0j, 0+0j, 0+0j, 0+0j, 0+0j, 0+0j, -0.132+0.088j, 0.302-0.125j, 0+0j]$^T$ |

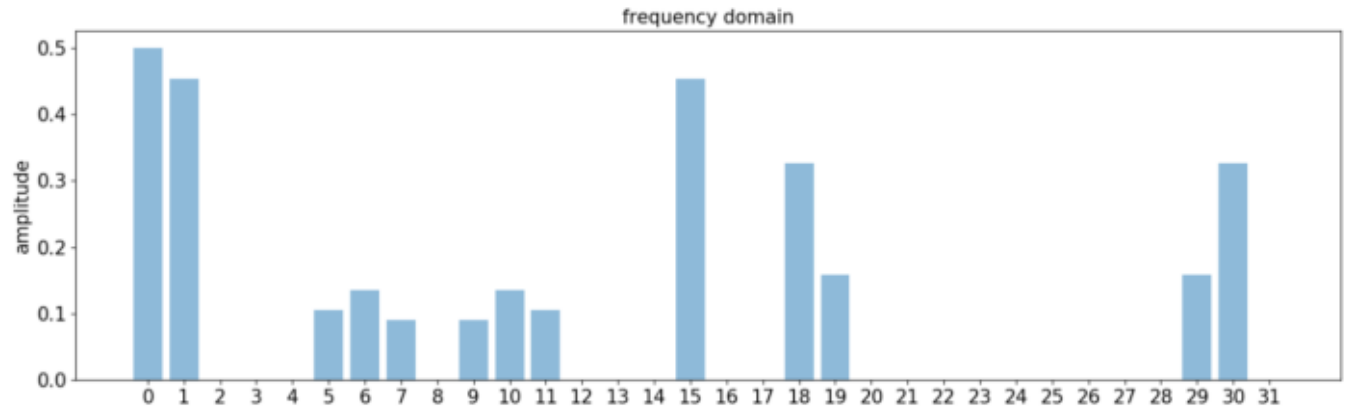

Fig. 49 Absolute Values after Filtering

| | |
|---|---|
| State vector after measurement (after 1 trial) | [0.583+0.j, -0.518-0.103j, 0+0j, 0+0j, 0+0j, 0.069+0.103j, -0.06, -0.146j, 0.02+0.103j, 0+0j, 0.02-0.103j, -0.06+0.146j, 0.069-0.103j, 0+0j, 0+0j, 0+0j, -0.518+0.103j, 0+0j, 0+0j, 0+0j, 0+0j, 0+0j, 0+0j, 0+0j, 0+0j, 0+0j, 0+0j, 0+0j ]$^T$ |

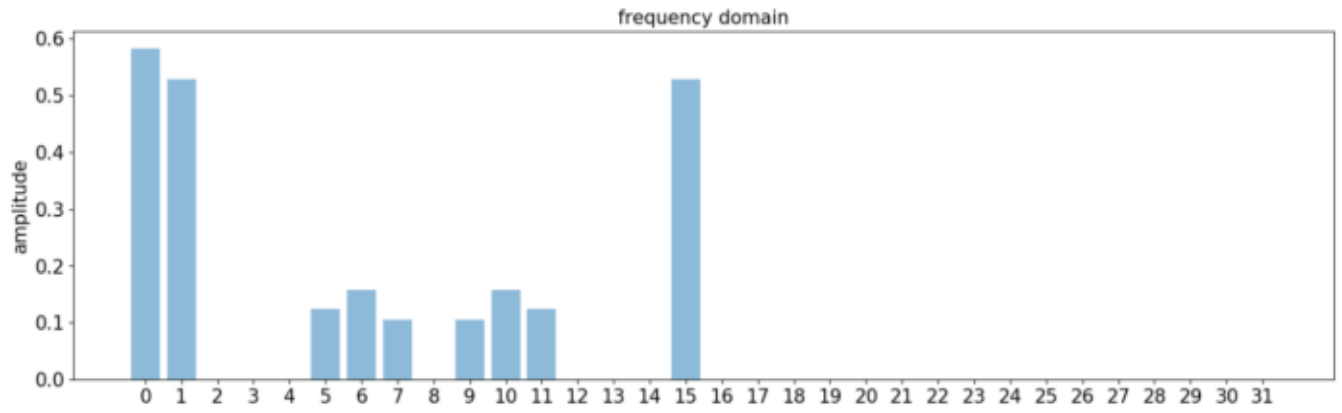

Fig. 50 Absolute Values after Measurement

| | |
|---|---|
| State vector after QFT | [-0.099+0j, -0.091+0j, -0.018+0j, 0.124+0j, 0.227+0j, 0.164+0j, 0.528-0j, 0.33 +0j, 0.33 +0j, 0.528+0j, 0.164+0j, 0.227+0j, 0.124+0j, -0.018+0j, -0.091+0j, -0.099+0j, 0,+0j, 0+0j, 0+0j, 0+0j, 0+0j, 0+0j, 0+0j 0+0j, 0,+0j, 0,+0j, 0,+0j, 0,+0j, 0+0j, 0+0j, 0+0j, 0+0j]$^T$ |

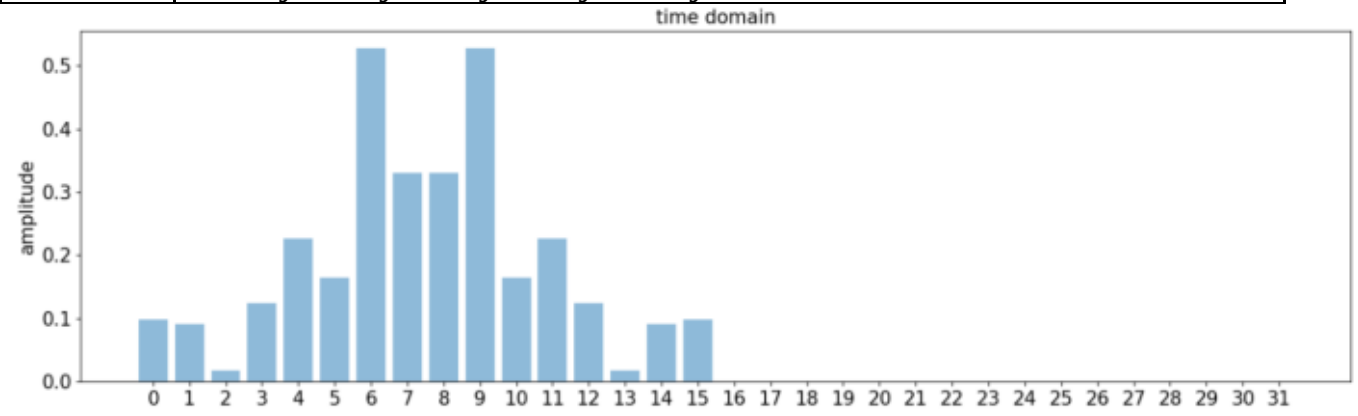

Fig. 51 Absolute Values after QFT

## MATLAB Band-Stop (Notch) Filtering

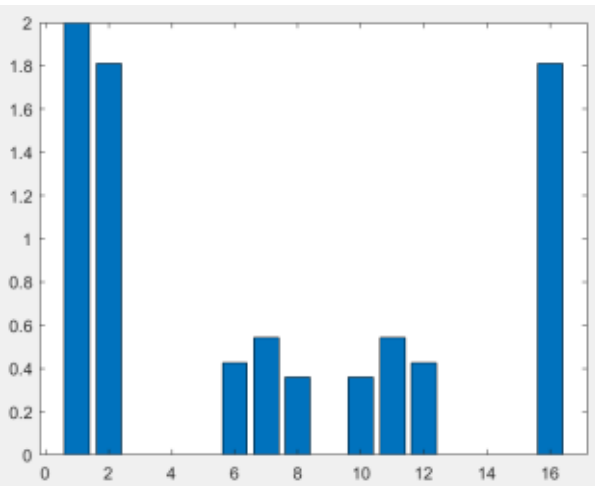

Fig. 52 Absolute Values after Filtering

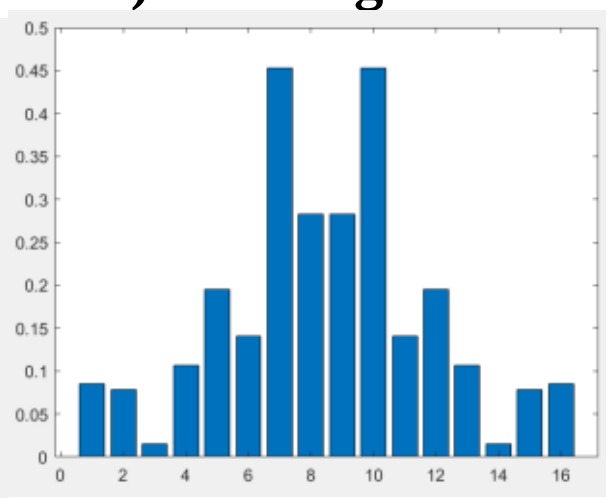

Fig. 53 Absolute Values after IFFT

| | |
|---|---|
| State vector after filtering | [2.0000+0.0000i, -1.7774-0.3536i, 0.0000+0.0000i, 0.0000+0.0000i, 0.0000+0.0000i, 0.2362+0.3536i, -0.2071-0.5000i, 0.0703+0.3536i, 0.0000+0.0000i, 0.0703-0.3536i, -0.2071+0.5000i, 0.2362-0.3536i, 0.0000+0.0000i, 0.0000+0.0000i, 0.0000+0.0000i, -1.7774 + 0.3536i]$^T$ |
| State vector after IFFT | [-0.0847, -0.0780, -0.0155, 0.1067, 0.1951, 0.1405, 0.4530, 0.2830, 0.2830, 0.4530, 0.1405, 0.1951, 0.1067, -0.0155, -0.0780, -0.0847]$^T$ |